\documentclass{emulateapj}
\usepackage{graphicx}
\usepackage{wasysym}

\begin{document}

\title{Is magnetic reconnection the cause of supersonic upflows in granular cells ?}

\author{J.M.~Borrero$^{1}$, V.~Mart{\'\i}nez Pillet$^{2,3}$, W.~Schmidt$^{1}$, C.~Quintero Noda$^{2,3}$, 
J.A.~Bonet$^{2,3}$, J.C.~del Toro Iniesta$^4$, L.R.~Bellot Rubio$^4$}
\affil{$^{1}$Kiepenheuer-Institut f\"ur Sonnenphysik, Sch\"oneckstr. 6, D-79110, Freiburg, Germany\\
$^{2}$Instituto de Astrof{\'\i}sica de Canarias, Avd. V{\'\i}a L\'actea s/n, La Laguna, Spain\\
$^{3}$Departamento de Astrof{\'\i}sica, Universidad de La Laguna, E-38205 La Laguna, Tenerife, Spain\\
$^{4}$Instituto de Astrof{\'\i}sica de Andaluc{\'\i}a (CSIC), Apdo. de Correos 3004, 18080 Granada, Spain}
\email{borrero@kis.uni-freiburg.de, vmp@ll.iac.es, wolfgang@kis.uni-freiburg.de, cqn@ll.iac.es, 
jab@ll.iac.es, jti@iaa.es, lbellot@iaa.es}

\begin{abstract}
{In a previous work, we reported on the discovery of supersonic magnetic upflows on granular cells in data from the {\sc Sunrise}/IMaX 
instrument. In the present work we investigate the physical origin of these events employing data of the same instrument but with higher 
spectral sampling. By means of the inversion of Stokes profiles we are able to recover the physical parameters (temperature, magnetic field, 
line-of-sight velocity, etc) present in the solar photosphere at the time of these events. The inversion is performed in a Monte-Carlo-like
fashion, that is, repeating it many times with different initializations and retaining only the best result. We find that many of the events 
are characterized by a reversal in the polarity of the magnetic field along the vertical direction in the photosphere, accompanied by an 
enhancement in the temperature and by supersonic line-of-sight velocities. In about half of the studied events, large blue-shifted and 
red-shifted line-of-sight velocities coexist above/below each other. These features can be explained in terms of magnetic reconnection, where 
the energy stored in the magnetic field is released in the form of kinetic and thermal energy when magnetic field lines of opposite polarities 
coalesce. However, the agreement with magnetic reconnection is not perfect and therefore, other possible physical mechanisms might also play a role.}
\end{abstract}


\shorttitle{Is magnetic reconnection the cause of supersonic upflows in granular cells ?}
\shortauthors{Borrero et al.}
\maketitle

\def\kms{~km s$^{-1}$}
\def\deg{^{\circ}}
\def\df{{\rm d}}

\section{Introduction}%

In the quiet solar photosphere the energy stored in the magnetic field is comparable to the kinetic energy due to convective motions.
This gives rise to a rich variety of phenomena that evolve at very short time and spatial scales. The IMaX instrument (Imaging 
Magnetograph eXperiment; Mart{\'\i}nez Pillet et al. 2011a ) onboard of stratospheric balloon {\sc Sunrise} (Barthol et al. 2011; Solanki et al. 2012) 
has helped to uncover many of these phenomena (see e.g. Solanki et al. 2010), from resolving magnetic 
flux-tubes (Lagg et al. 2010), to finding vortex tubes (Steiner et al. 2010), vortex flows (Bonet et al. 2010), etc. One of those discoveries 
involves supersonic magnetic upflows (Borrero et al. 2010, 2012). These events are characterized by highly blue-shifted circular 
polarization signals, that appear at the center or edges of granular cells and last for about 80 seconds. Similar 
events were subsequently found also in data from the SP instrument onboard the Hinode spacecraft (Mart{\'\i}nez Pillet et al. 2011b).
The fact that magnetic fields of opposite polarity connected by horizontal fields appear in the vicinity of these 
events in 70 \% of the cases, led us to surmise that they are caused by magnetic reconnection. In this paper we study them 
in more detail using a different data set from the IMaX instrument. A comparison between the old and new data sets is provided in
Section 2. In Section 3 we describe the criterion employed to select events.
The analysis technique, namely, the inversion of the observed Stokes profiles to retrieve the physical conditions of the solar atmosphere, is 
detailed in Section 4. Section 5 presents our results while Section 6 briefly addresses the choice of model for the inversion. Finally,
Section 7 presents our main conclusions.\\

\section{Instruments and observations}%

The data employed in this work were recorded with the stratospheric balloon-borne observatory {\sc Sunrise} (Solanki et al. 2010; 
Barthol et al. 2011). {\sc Sunrise} was launched on June 8, 2009 from Kiruna (Sweden) and landed on June 13, 2009 on Somerset 
Island (Canada). During this time, {\sc Sunrise}'s 1-meter telescope took broad-band images in different spectral windows with 
the SUFI instrument (Gandorfer et al. 2011), and spectropolarimetric data of the solar photosphere with IMaX (Mart{\'\i}nez Pillet et al. 2011a).
 An average flight altitude of 35 km allowed {\sc Sunrise} to avoid more than 95 \% of the disturbances introduced by Earth's atmosphere.
In addition, image motions due to wind during the flight were stabilized by the Correlation-Tracker and Wavefront Sensor 
(CWS; Berkefeld et al. 2011). Owing to the aforementioned advantages, IMaX spectropolarimetric data yielded a spatial 
resolution of 0.25" and a field-of-view of 50"$\times$50". Further image reconstruction based on phase diversity calibration of
the PSF of the optical system improved the resolution to 0.15"-0.18".\\

In Borrero et al. (2010) we employed reconstructed IMaX data that included the four components of the Stokes vector
($I$, $Q$, $U$, $V$) measured at five wavelength positions across the \ion{Fe}{1} 5250.217 {\AA} spectral line. In the following
we will refer to this observing mode as V5-6. In this work, however, we will use a different observing mode, 
referred to as L12-2. In this mode the intensity $I$ and circular polarization $V$ were measured in twelve (instead of five) 
wavelength positions). For reasons that will be explained later, we restricted ourselves to 
employ non-reconstructed data with a spatial resolution of \emph{only} 0.25".\\

Figure 1 displays the region of the intensity spectra around the \ion{Fe}{1} 5250.217 spectral line, as recorded by the 
Fourier Transform spectrometer in the quiet Sun (Wallace et al. 1998). Crosses in this figure show the five wavelength
positions scanned by the V5-6 data used in Borrero et al. (2010). These were located at $\Delta  = 
[-80,-40,40,80,227]$ m{\AA} from line-center. Filled circles illustrate the wavelength positions 
scanned in the L12-2 observing mode. Here, the wavelength range goes from $-192.5$ m{\AA} to $+192.5$ m{\AA}, in 
twelve positions equidistantly distributed in steps of 35 m{\AA}.\\

\begin{center}
\includegraphics[width=9cm]{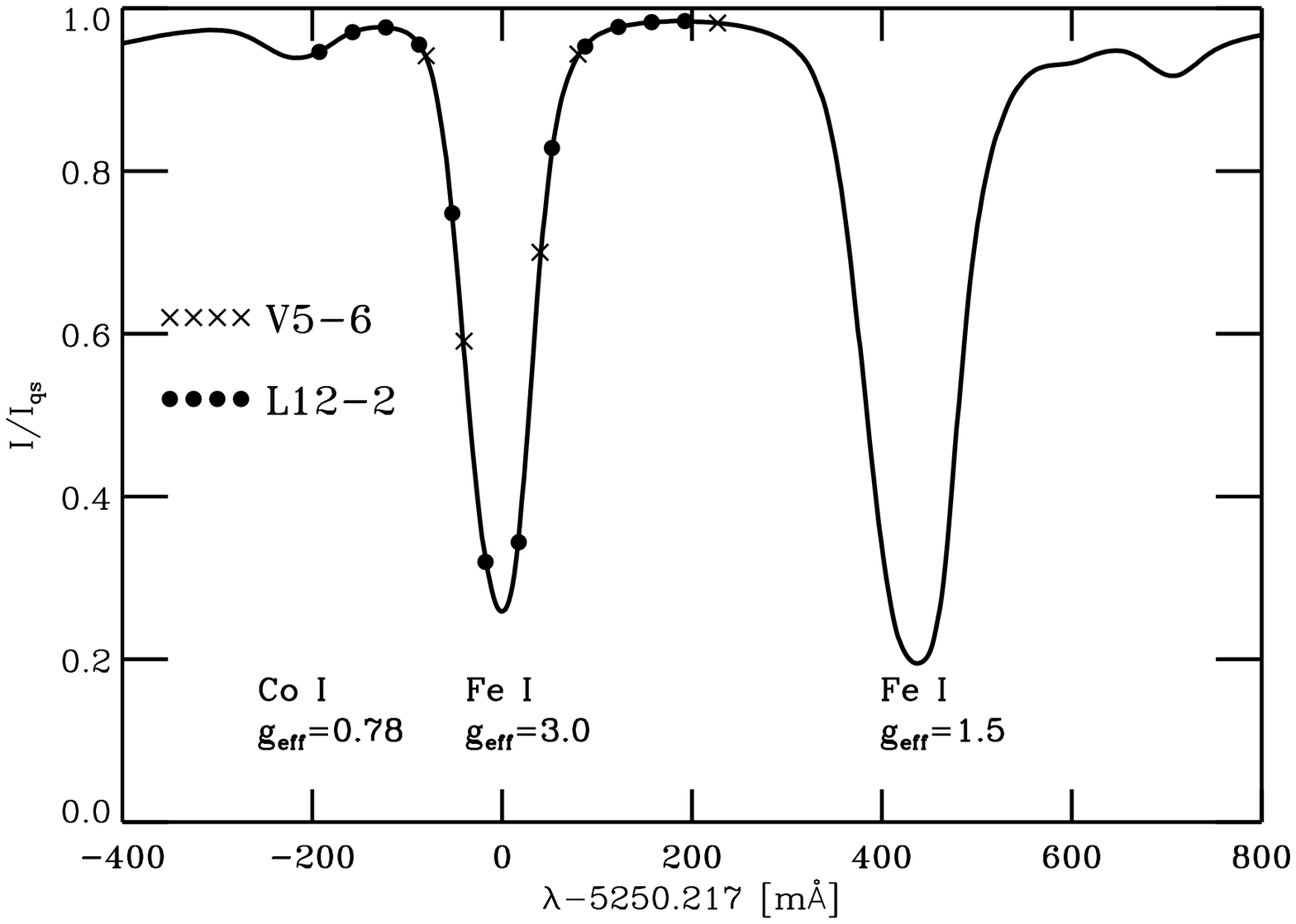}
\figcaption{Comparison of the V5-6 (crosses) and L12-2 (filled circles) observing modes. The former mode records
the four components of the Stokes vector, while the latter acquires Stokes $I$ and Stokes $V$ (see text for details). The solid-black
line corresponds to the Fourier Transform Spectrometer data (FTS-atlas) from Kitt Peak observatory in Arizona. The effective
Land\'e factors $g_{\rm eff}$ of each line are calculated under the LS approximation from the electronic configurations given in Table 1.}
\end{center}

The lack of linear polarization profiles $Q$ and $U$ in the L12-2 observing mode makes the polarimetric calibration
of the data slightly more difficult to implement compared to V5-6. The instrument's calibration matrix had been theoretically 
calculated (Mart{\'\i}nez Pillet et al. 2011a) and experimentally confirmed at the INTA (Instituto Nacional de T\'ecnicas 
Aeroespaciales) facilities in Spain (Mart{\'\i}nez Pillet 2007; del Toro Iniesta \& Mart{\'\i}nez Pillet 2012) such that
linear polarization cross-talk could be minimized by tuning the voltages of the nematic liquid crystals to the appropiate
retardances. Unfortunately, this calibration stage was not performed while mated to the Sunrise telescope (main and secondary
mirrors) which introduced more cross-talk than anticipated. Therefore, we expected to find a non-negligible contribution 
from $Q$ and $U$ in the measured linear combinations of $I$ and $V$. Fortunately, the events that we will focus on, 
the so-called {\it supersonic magnetic upflows in granular cells}, had already been analyzed using the V5-6
observing mode. From that data we learned that these events have a negligible amount of linear polarization 
(although patches of enhanced linear polarization usually appear within 2" of these events; see for instance Fig.~4 in Borrero
et al. 2010 and Figs.~3-7 in Borrero et al. 2012). This means that, even in our scenario, where there is a significant cross-talk
 from $Q$ and $U$ into Stokes $V$, the circular polarization is not affected much. The uncorrected cross-talk does however
slightly increase the noise level in the circular polarization.\\

IMaX used the L12-2 mode to observe the solar photosphere for about 60 minutes on June 10, 2009. The observations
were recorded close to disk center, $\mu = \cos\Theta = 0.99$, where $\Theta$ corresponds to the heliocentric angle.
Due to pointing problems, the time series was interrupted several times and only about half of that time is usable. In total, there are 
52 full scans of Stokes $I$ and $V$ in twelve wavelength positions. Each scan is recorded during a time interval 
of 33 seconds. The noise-level is estimated to be $10^{-3}$ in units of the average quiet Sun continuum intensity.\\

\section{Selection of Events}%

In Borrero et al. (2010), where we employed V5-6, the supersonic magnetic upflows were 
detected in the circular polarization (Stokes $V$) at $\Delta \lambda = + 227$ m{\AA} (see Fig.~1). 
At this wavelength any signal can be produced by either a strong red-shift (downflow) from 
\ion{Fe}{1} 5250.217 {\AA} or by strong blue-shift (upflow) from \ion{Fe}{1} 5250.653 {\AA}. In 
the aforementioned work, the signal was ascribed to the latter case because the 
Stokes $I$ signal from \ion{Fe}{1} 5250.217 {\AA} was blue-shifted.\\

Bearing this information in mind, we look for a strategy to find highly blue-shifted Stokes $V$ profiles in the
L12-2 data. The first idea that comes to mind is to use the blue-most wavelength position in 
these data, $\Delta \lambda = -192.5$ m{\AA}, to find such upflows. Unfortunately, this scanning position is 
not far enough towards the blue to guarantee that the observed signal at this wavelength is only due to large 
velocities, as it could also be caused by a magnetic field that shifts the $\sigma$-component of Stokes $V$ into
this wavelength. Indeed, displaying $V(\lambda = \lambda_0-192.5~ \textrm{m}${\AA}$)$, reveals a pattern that closely resembles
the network. Thus, this wavelength alone cannot be employed to uniquely identify large upflows.\\

In order to disentangle network elements from possible supersonic magnetic upflows, we propose a different
strategy, in which we compare the circular polarization close to the spectral line core and the circular polarization
close to the blue continuum. Let us refer to these two quantities as $V_{\rm c}$ and $V_{\rm line}$, respectively. They
are defined as:

\begin{eqnarray}
V_{\rm line} = \frac{1}{3} \left[\sum_{i=4}^{i=6} V(\lambda_i) - \sum_{i=7}^{i=9} V(\lambda_i)\right]\\
V_{\rm c} = \frac{1}{2} \sum_{i=1}^{i=2} |V(\lambda_i)| \;,
\end{eqnarray}

\noindent where the index $i$ runs from the blue-most ($i=1$) o the red-most ($i=12$) scanning positions indicated by the filled circles
in Fig.~1. We note that, in the definition of $V_{\rm line}$ we are subtracting Stokes $V$ in the red wing ($i=6,7,8$) 
from Stokes $V$ in the blue wing ($i=3,4,5$). This is done in order to obtain the polarity of the magnetic field vector,
and has the additional benefit of partially canceling the noise.\\

In the top-left panel of Figure 2 we display $V_{\rm line}$ (normalized to the averaged quiet Sun continuum intensity 
$I_{\rm qs}$) over a portion of the full field-of-view from one of our available 52 snapshots. The regions
of enhanced $V_{\rm line}$ correspond mostly to the network elements since the circular polarization close to the line-center is
large. In the top-right panel of Figure 2 we plot, for the same region, the absolute value of the quotient of 
$V_{\rm c}$ and $V_{\rm line}$. In this panel, the network appears as those regions where 
$\| V_{\rm c} / V_{\rm line}\| \rightarrow 0$. Regions where $\| V_{\rm c} / V_{\rm line}\| >> 1$
denote Stokes $V$ profiles that are highly blue-shifted. Our selection criterion will consider as {\it supersonic
magnetic upflows} any pixel in the field-of-view where $\| V_{\rm c} / V_{\rm line}\| > 4$. In Figure 2, those
regions are indicated by the white contours. Note that the selected events also coincide with the center/edges
of granular cells where the line-of-sight velocity is blue-shifted by about $-2$ km s$^{-1}$ (see black contours
in the left-bottom and right-bottom panels in Figure 2). This confirms that the selected events have the same properties
as those studied in Borrero et al. (2010).\\

The line-of-sight velocity $V_{\rm LOS}$ displayed in Figure 2 is obtained by calculating 
the center-of-gravity of Stokes $I$ and correcting for gravitational red-shift, convective blue-shift, and for the 
wavelength shift across the FOV due to the collimated configuration of the instrument (see Sect.~9.1 in Mart{\'\i}nez Pillet 
et al. 2011a). We mention this in order to avoid confusion with the $V_{\rm LOS}$ that will be used in the next sections
of this paper, which will be inferred from the simultaneous fitting of Stokes $I$ and $V$.\\

It is also important to clarify that Figure 2 shows a somewhat exceptional situation, in which three events occur 
on a small portion of the full field-of-view. This case has been selected to highlight the properties of the selected
events, but by no means corresponds to the typical case seen in the observations. In fact, from the 52 available snapshots,
the aforementioned selection criteria selects 857 pixels, belonging to 122 events. This results in an average of 2.3 events
in each 50"$\times$50" snapshot. Unfortunately, the pointing problems described in Sect.~2 prevented us from having 
a continuous time-series and therefore we cannot track each event in time. Thus, some of those 122 events might
correspond to the same one but at different times in their evolution. Consequently, we cannot 
compare these numbers with the occurrence rates obtained in Borrero et al. 2010.\\

\begin{figure*}
\begin{center}
\begin{tabular}{cc}
\includegraphics[width=9cm]{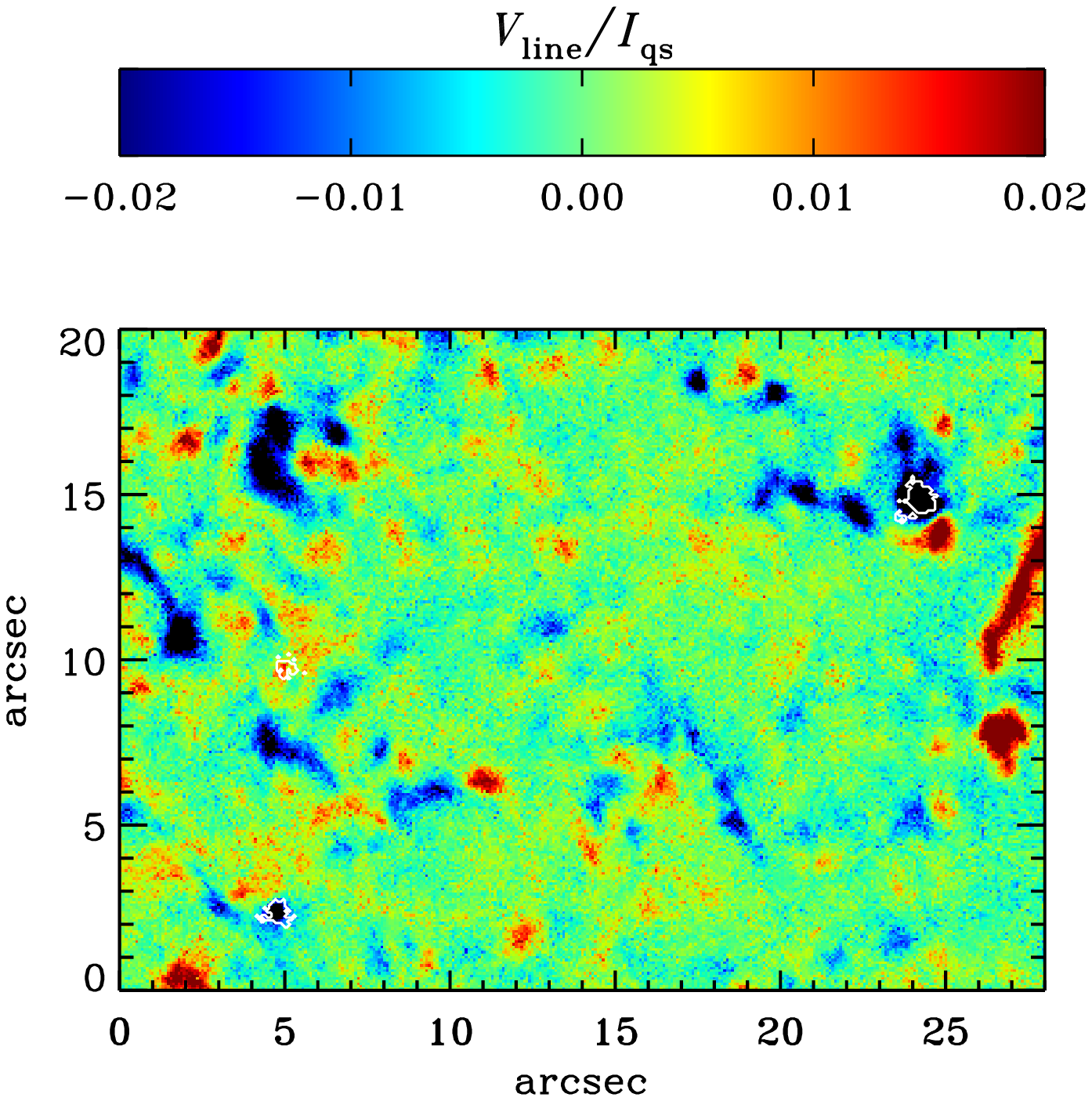} &
\includegraphics[width=9cm]{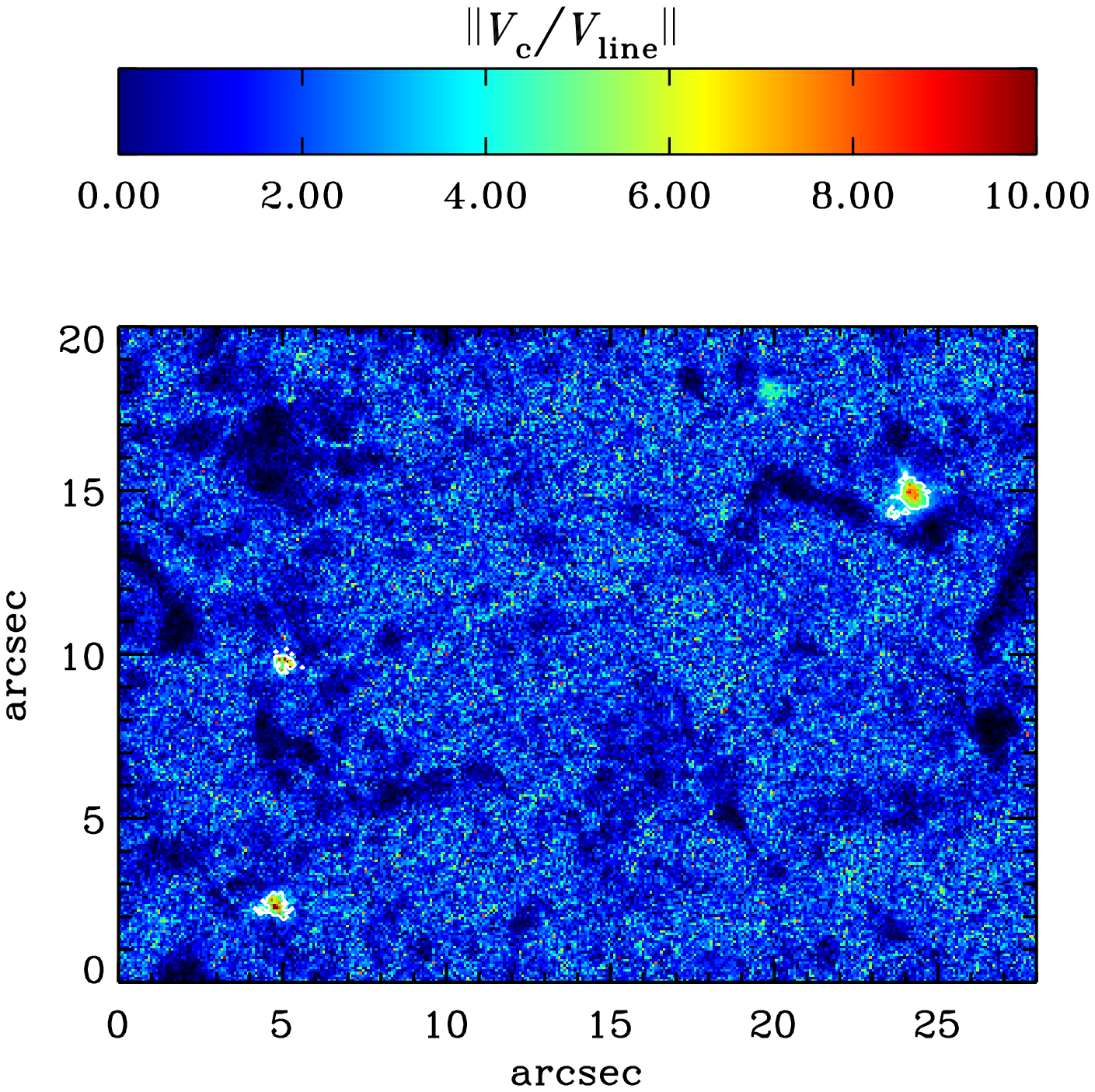} \\
\includegraphics[width=9cm]{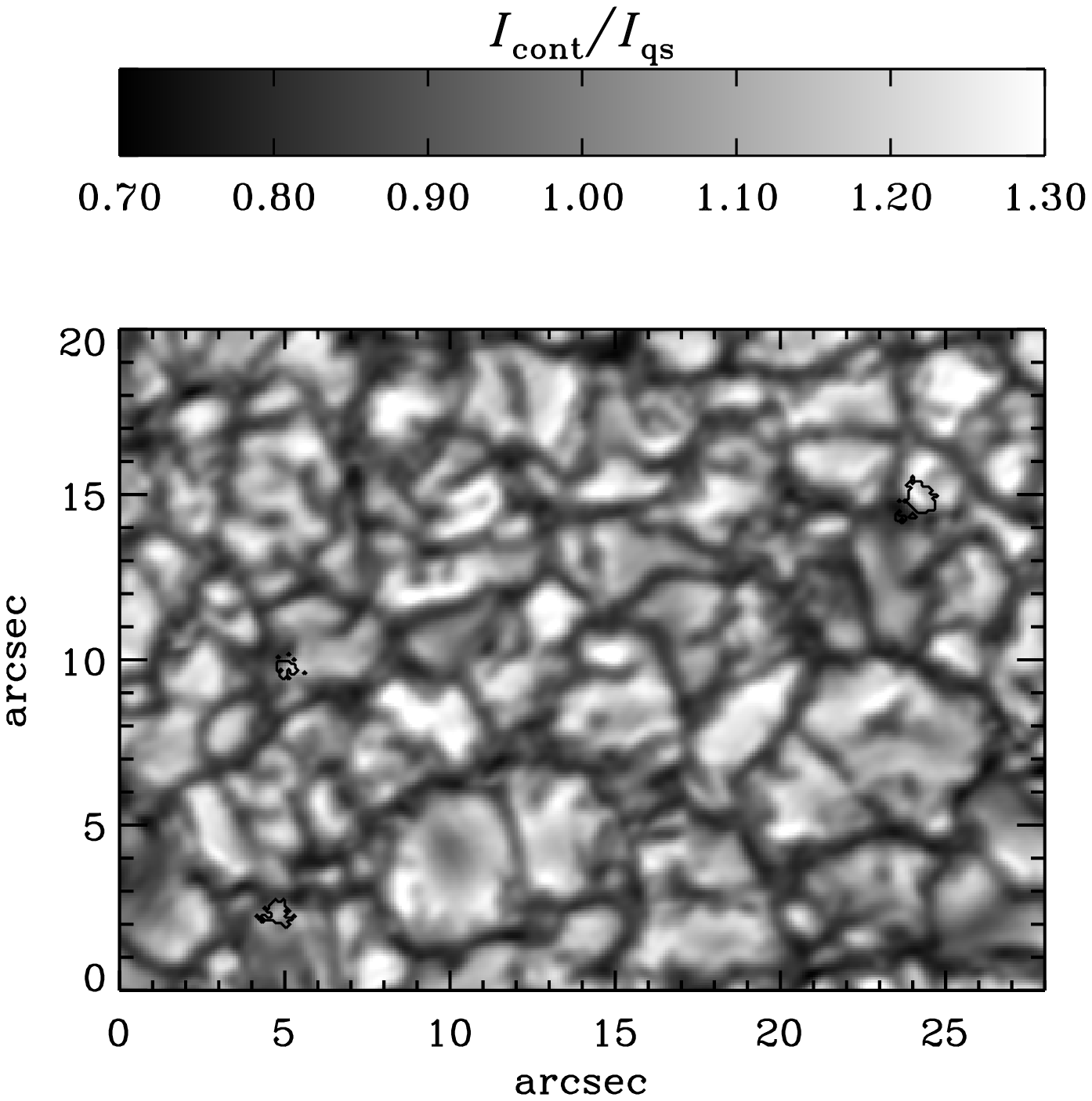} &
\includegraphics[width=9cm]{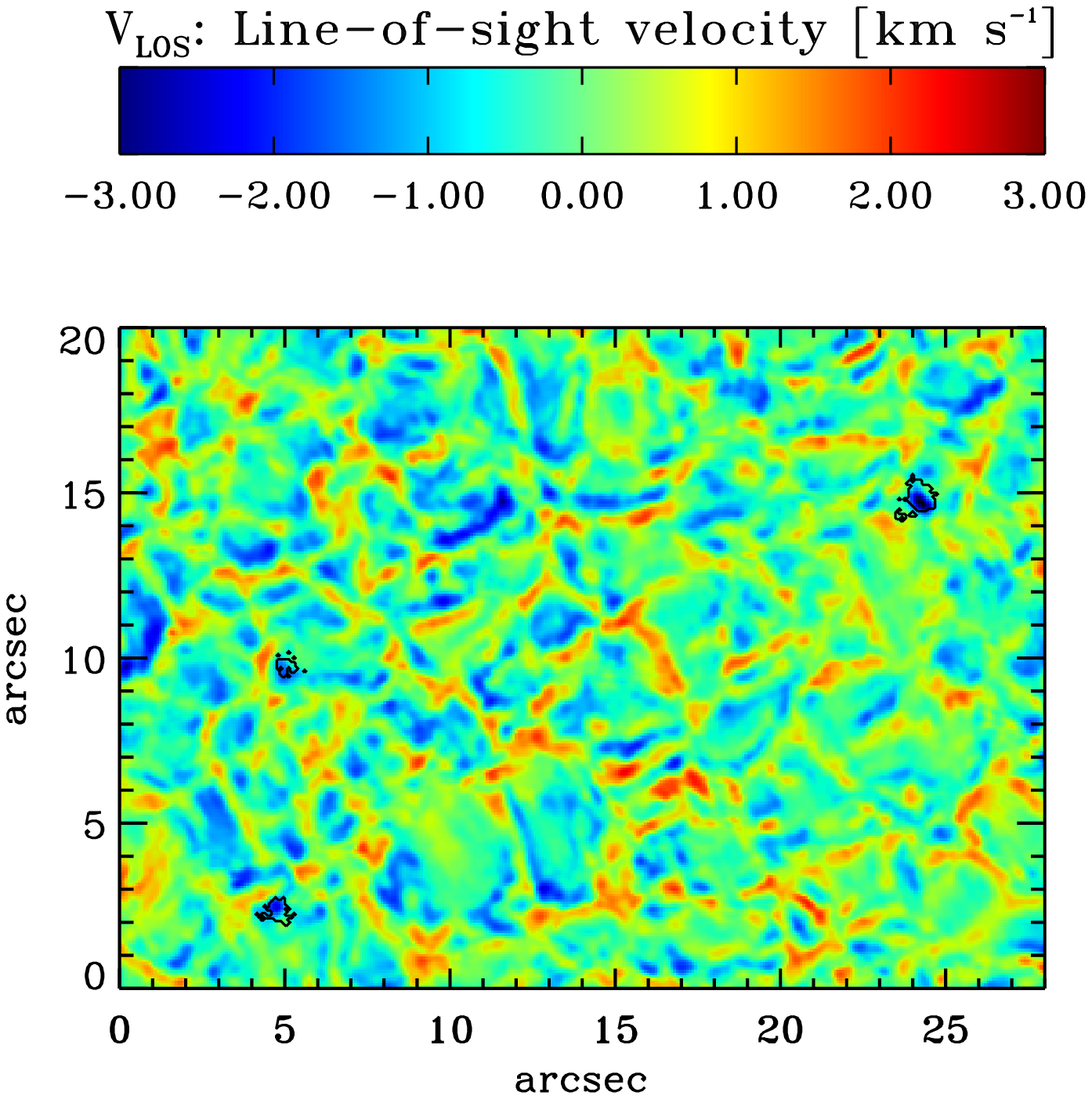} \\
\end{tabular}
\figcaption{{\it Top-left}: close-up ($28"\times 25"$) of a snapshot displaying the total circular polarization 
around the center of the spectral line $V_{\rm line}$ (Eq.~1) and normalized to the average continuum intensity over the
quiet Sun $I_{\rm qs}$. {\it Top-right}: same FOV as before but showing $\| V_{\rm c} / V_{\rm line}\|$, where $V_{\rm c}$ 
corresponds to the circular polarization on the blue-wing of the spectral line (Eq.~2). {\it Bottom-left}: same FOV 
as before but showing the continuum intensity $I_{\rm c}$ normalized to the average continuum intensity over the quiet Sun
$I_{\rm qs}$. {\it Bottom-right}: line-of sight velocity $V_{\rm LOS}$ derived from the center-of-gravity of Stokes $V$. In
all panels the black and white contours enclose the regions where $\| V_{\rm c} / V_{\rm line}\| > 4$. These patches contain 
pixels that are selected for our study (see text for details). While the upper panels are unreconstructed, the bottom ones 
have been subject, for visualization purposes, to image restoration (see Sect.~2).}
\end{center}
\end{figure*}

\section{Inversion of Stokes profiles}%

Once we have the selected pixels that correspond to the possible {\it supersonic magnetic upflows in granules},
we now proceed to extract physical information from their corresponding Stokes $I$ and $V$ profiles. This is done
by means of the inversion of the radiative transfer equation employing the SIR (Stokes Inversion based on Response functions)
code (Ruiz Cobo \& del Toro Iniesta 1992). Starting with an initial model of the solar photosphere, SIR solves the
radiative transfer equation to obtain the theoretical Stokes vector that arises from such model. The observed Stokes vector is then
compared to the theoretical one through a $\chi^2$ merit-function. Via a Levenberg-Marquardt method, the original model is then iteratively 
modified until $\chi^2$ reaches a minimum. At each iteration step, the Levenberg-Marquardt algorithm provides
the perturbations in the physical parameters at several optical-depth positions called {\it nodes}, that are needed to produce a 
better fit to the observed Stokes profiles. Each node represents a free parameter in the inversion.
The resulting model (from the $\chi^2$-minimization) can be then considered to represent the physical conditions present in 
the solar photosphere. For some recent reviews on this subject we refer the reader to del Toro Iniesta (2002), Bellot Rubio (2006) 
and Ruiz Cobo (2007). Since photon noise affects the results of the inversion rather negatively (see e.g. Borrero \& Kobel 2011, 2012) 
and owing to the fact that image reconstruction techniques slightly increase the noise in the observations, we considered for 
the inversion the unreconstructed data with a spatial resolution of 0.25 arcsec (see Sect.~2).\\

Our inversions have been carried out with a 1-component model, in which the photosphere is considered 
to be laterally homogeneous within each selected pixel. Thus, we only need to consider
the vertical variations of the physical parameters in the photosphere. These variations are often
treated in terms of the dimensionless optical-depth at a reference wavelength of 5000 {\AA}, $\tau_5$, instead of the
geometrical height $z$. In general, the physical parameters relevant for the formation of spectral lines are: temperature $T(\tau_5)$,
line-of-sight velocity $V_{\rm LOS}(\tau_5)$, and the three components of the magnetic field vector: $B(\tau_5)$ (modulus
of the magnetic field vector), $\gamma(\tau_5)$ (inclination of the magnetic field vector with respect to the observer's
line-of-sight) and $\phi(\tau_5)$ (azimuth of the magnetic field vector in the plane perpendicular to the observer's
line-of-sight). Other quantities, such as the gas $P_{\rm g}(\tau_5)$ and electron $P_{\rm e}(\tau_5)$ pressure, as well
as the density $\rho(\tau_5)$, are derived from $T(\tau_5)$, the condition of hydrostatic equilibrium, and the equation of ideal
gases with a variable mean molecular weight. In our inverions we allow for the following free parameters (nodes): three for 
$T(\tau_5)$, one for $B(\tau_5)$ (constant value with height), two for $\gamma(\tau_5)$, five for $V_{\rm LOS}(\tau_5)$, and 
finally one for the micro-turbulent velocity $V_{\rm mic}(\tau_5)$ (also constant with height). This adds up to a total of 12 
free parameters. The full stratification of the physical parameters with $\tau_5$ is obtained via interpolation across the 
values at the nodes. In section 6 we give more details about our choice of model and free parameters, as well as discussing its 
implications.\\

\begin{center}
\includegraphics[width=9cm]{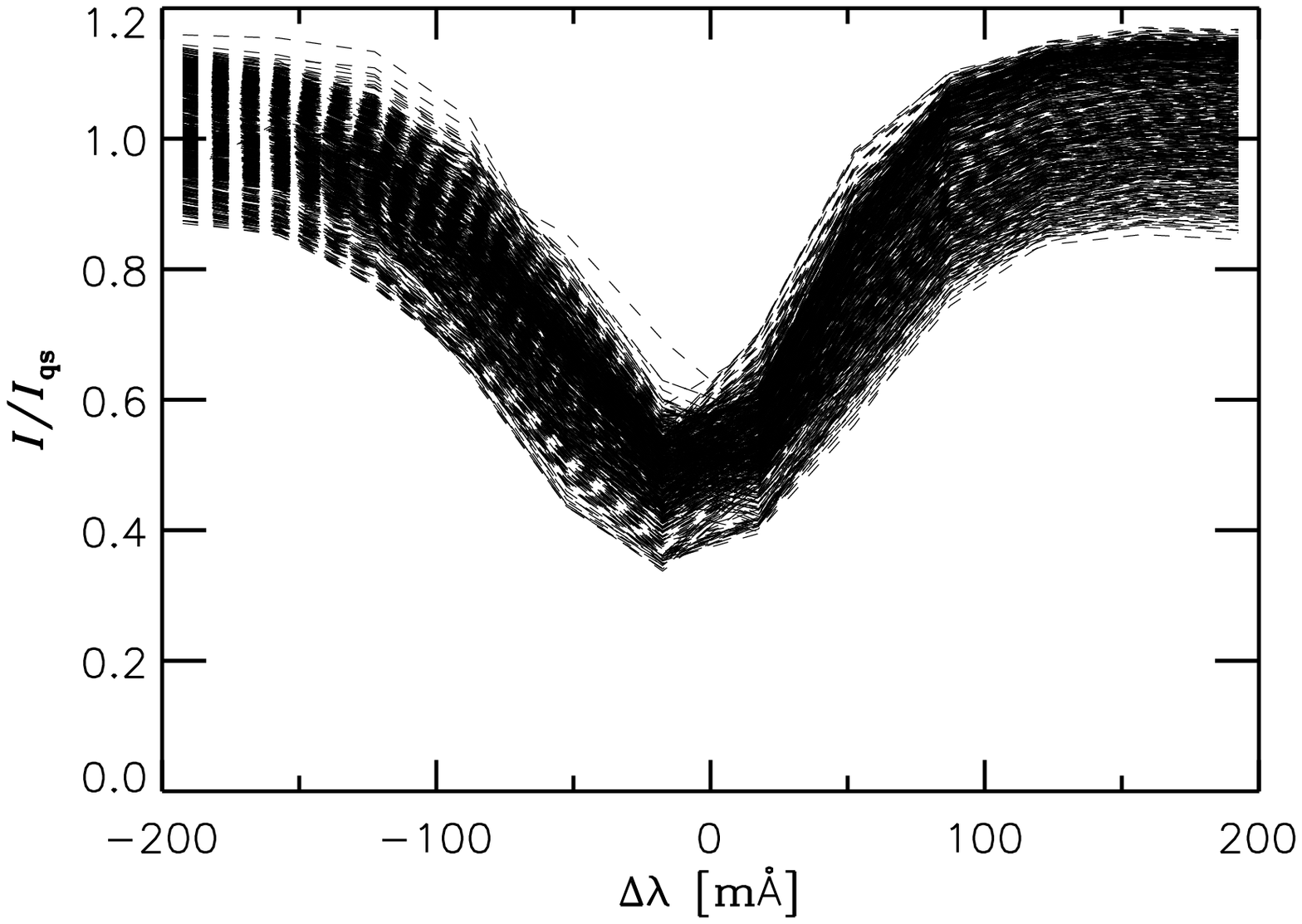}\\
\includegraphics[width=9cm]{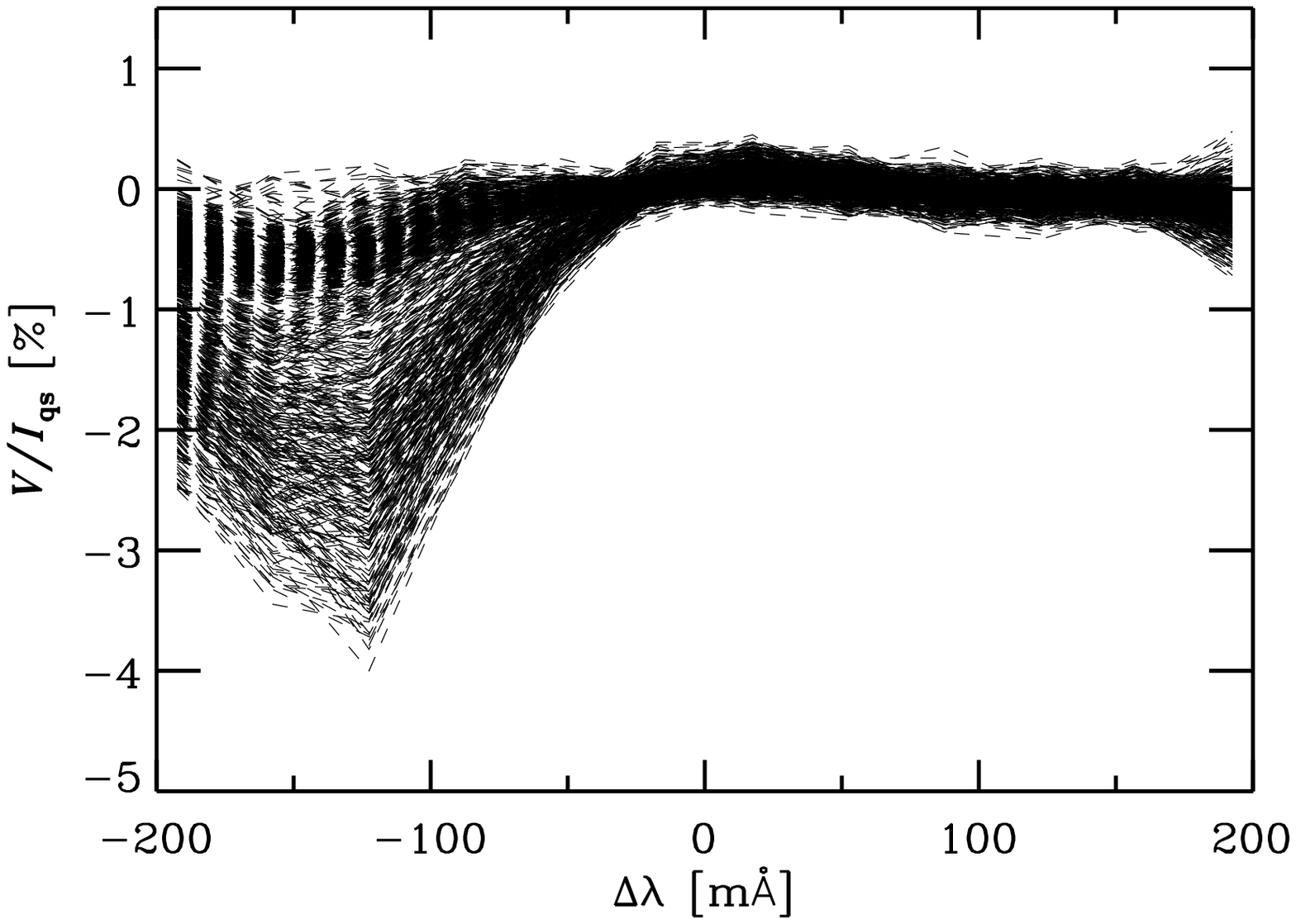}\\
\figcaption{Observed Stokes $I$ (upper panel) and Stokes $V$ (bottom panel) profiles from the 857 selected pixels. 
For better visualization we have interpolated the data to a common wavelength grid and we have changed the sign 
in Stokes $V$ so that the of lobe is always negative. All profiles are normalized to the average continuum intensity 
of the quiet Sun $I_{\rm qs}$.}
\end{center}

It is important to remind here that the L12-2 observing mode does not record the linear polarization 
profiles ($Q$, $U$; see Sect.~2). In the case of strong magnetic fields it is plausible to recover, through 
the magneto-optical effects, the azimuth of the magnetic field vector $\phi$ from only Stokes $I$ and $V$ 
(Ruiz Cobo \& del Toro Iniesta 1992). However, it is unclear whether this is also the case when the 
circular polarization signals are weak ($V/I_{\rm qs} \leq 0.03$; see Fig.~1 and also Borrero \& Kobel 2011), 
and the spectral resolution is limited. Therefore, we do not invert the azimuth angle of the magnetic field vector 
$\phi(\tau_5)$, and instead we fix it at a value of zero. Finally, we note that IMaX's spectral transmission profile 
is fully considered in the inversion. This is done by convolving the theoretical Stokes vector, before it is compared 
to the observed Stokes vector at each iteration step, with the instrument's transmission curve (see Fig.~2 in
Mart{\'\i}nez Pillet et al. 2011a).\\

In order to improve convergence, and to reduce the chances of the Levenberg-Marquardt algorithm falling into a local 
minimum, each of the 857 selected pixels is inverted a total of 100 times, where the initial values of the physical parameters 
are randomly chosen each time. From all those 100 independent inversions we retain only the one with the smallest value of $\chi^2$. 
In the inversion we include the effects of the three spectral lines present in Figure 1. This is done because 
it is not possible to rule out the possibility of the \ion{Co}{1} 5250.008 {\AA} and/or \ion{Fe}{1} 5250.653 {\AA} spectral 
lines entering the wavelength regions scanned by the L12-2 observing-mode (see filled circles in Figure 1), 
specially when considering events that involve large line-of-sight velocities\footnote{In fact, as mentioned in the 
first paragraph in Sect.~3, this already happened in the V5-6 data (see Borrero et al. 2010 for details).}. The atomic parameters 
for these three spectral lines are given in Table 1.\\

\begin{deluxetable}{cccccccc}
\tablecaption{Atomic parameters of the spectral lines included in the inversion.}
\tablehead{\colhead{Specie} & \colhead{$\lambda_{\astrosun}$\tablenotemark{1}} & \colhead{$\chi_{\rm low}$} & \colhead{$\log(gf)$} 
& \colhead{Elec.conf} & \colhead{$\sigma$}\tablenotemark{2} & \colhead{$\alpha$}\tablenotemark{2} & \colhead{$g_{\rm eff}$}\\
\colhead{} & \colhead{[{\AA}]} & \colhead{[eV]} & \colhead{} & \colhead{} & \colhead{} & \colhead{} & \colhead{}}
\startdata
\ion{Co}{1} & 5250.008 & 4.175 & $-$0.114 & ${^4}G_{5/2}-{^4}H_{7/2}$ & n/a & n/a & 0.785 \\
\ion{Fe}{1} & 5250.217 & 0.121 & $-$4.938 & ${^5}D_0-{^7}D_1$ & 207 & 0.253 & 3.0\\
\ion{Fe}{1} & 5250.653 & 2.198 & $-$2.198 & ${^5}P_2-{^5}P_3$ & 344 & 0.268 & 1.5\\
\enddata
\tablenotetext{1}{$\lambda_{\astrosun}$ represents the central wavelength position of the spectral line on the Sun.}
\tablenotetext{2}{$\sigma$ and $\alpha$ represent the atomic transition's cross-section (in units of Bohr's 
radius squared $a_0^2$) and velocity parameter, respectively, for collisions with neutral atoms under the ABO theory (Asntee \& 
O'Mara 1995; Barklem et al. 1998). Collisional data for \ion{Co}{1} is not available, and therefore 
we rely on Uns\"old's theory (Uns\"old 1955) to calculate the collisional broadening of the spectral line.}
\end{deluxetable}

\section{Inversion results}%

The inversion of the Stokes $I$ and $V$ profiles from the 857 selected profiles provides (among other physical parameters) 
the stratification with optical depth of the temperature $T(\tau_5)$, line-of-sight velocity $V_{\rm LOS}(\tau_5)$, magnetic 
field strength $B(\tau_5)$, and inclination $\gamma(\tau_5)$ in all selected pixels. Searching for similarities among all 
available results turns out to be a difficult task, as the inferred stratifications do not seem to follow, at first glance, 
any particular pattern. However, if one looks at the optical-depth dependence of the line-of-sight component of the magnetic 
field vector $B_\parallel = B \cos \gamma$, one realizes that almost all pixels show a change in sign from $B_{\parallel} > 0$ 
to $B_{\parallel} < 0$ (or viceversa) at some optical-depth point $\tau_5$ (height in the atmosphere). This observation allows 
us to classify the different results as a function of the optical-depth where the polarity of the magnetic 
field reverses. In particular we distinguish three cases: polarity change at around $\log\tau_5 \approx -1$, $\log\tau_5 \approx -2$, 
and finally, possible polarity change at $\log\tau_5 < -3$. Hereafter, these families of solutions will be referred 
to as {\it family 1}, {\it family 2}, and {\it family 3}, respectively. Figures 4, 5 and 6 show the 
individual results as a function of $\tau_5$ from the inversion of all pixels belonging to each family (black-dashed lines). 
Each of these figures show: the line-of-sight component of the magnetic field vector $B_\parallel$ (top-left panel), the temperature 
$T$ (top-right panel), and the line-of-sight velocity $V_{\rm LOS}$ (bottom-left panel). For the latter two physical parameters, $T(\tau_5)$ and 
$V_{\rm LOS}(\tau_5)$, we also show in solid-red lines the average stratification obtained from all the pixels belonging to a given family. 
For comparison purposes, the top-right panels in Figs.4, 5 and 6 also display the average temperature stratification (blue-solid 
lines) in granules (Borrero \& Bellot Rubio 2002). This model is chosen because, as mentioned in Sect.~3, these event occurs 
typically at the center or edges of granular cells (see also Fig.~2). In the following, we will describe each of the aforementioned 
families separately.\\
\subsection{Family 1: polarity change at $\log\tau_5 \approx -1$.}

This family comprises 123 pixels out of the 857 selected ones (14.3 \%). As imposed by our classification 
criterion, $B_\parallel$ changes sign at around $\log\tau_5 \approx -1$ (Fig.~4; top-left panel). The temperature shows an enhancement 
of about $400-600$ K in the mid- and upper-photosphere ($\log\tau_5 \in [-1.5,-3]$) with respect to a typical granule 
(Fig.~4; top-right panel). The line-of-sight velocity (Fig.~4; bottom-left panel) displays variations from extreme downflows
($V_{\rm LOS} \approx 12$\kms) in the upper-photosphere ($\log\tau_5 \approx -3$) to large upflows in the mid-photosphere
($V_{\rm LOS} \approx -7$\kms~ at $\log\tau_5 \approx -2$), and then back to downflows in the deep-Photosphere ($V_{\rm LOS} 
\approx 3$\kms~ at $\log\tau_5 \approx 0$). Since the speed of sound in the solar photosphere is about $V_s \simeq 7$\kms, the inferred 
line-of-sight velocities are close to supersonic. Once we consider that $V_{\rm LOS}$ is only a lower limit of the total modulus of the 
velocity vector, the final velocities are likely to be much larger, hence supersonic. An additional feature is the fact that the line-of-sight 
velocity remains close to zero where the polarity of the magnetic field changes ($\log\tau_5 \approx -1$).\\

\begin{figure*}
\begin{center}
\begin{tabular}{cc}
\includegraphics[width=9cm]{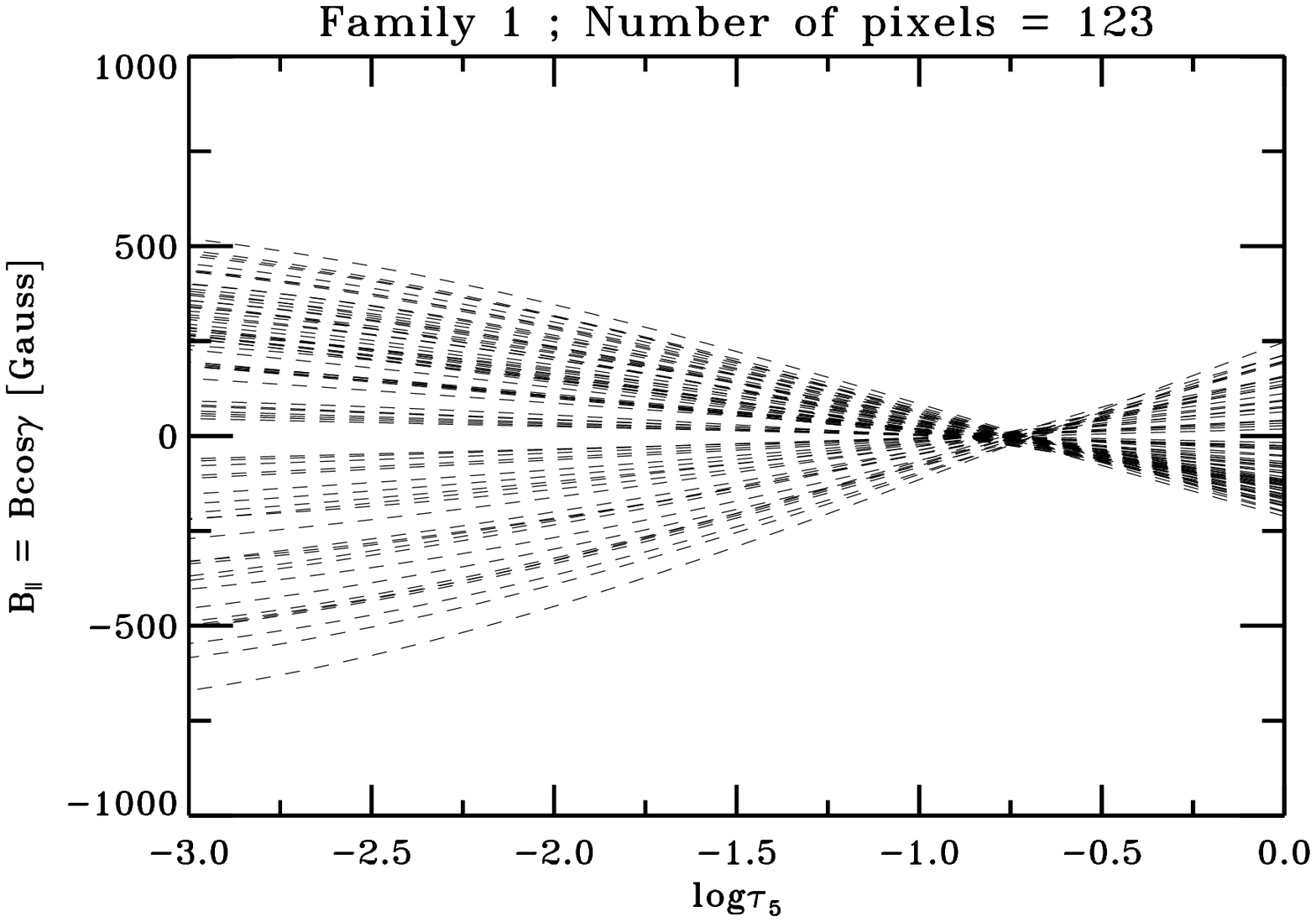} &
\includegraphics[width=9cm]{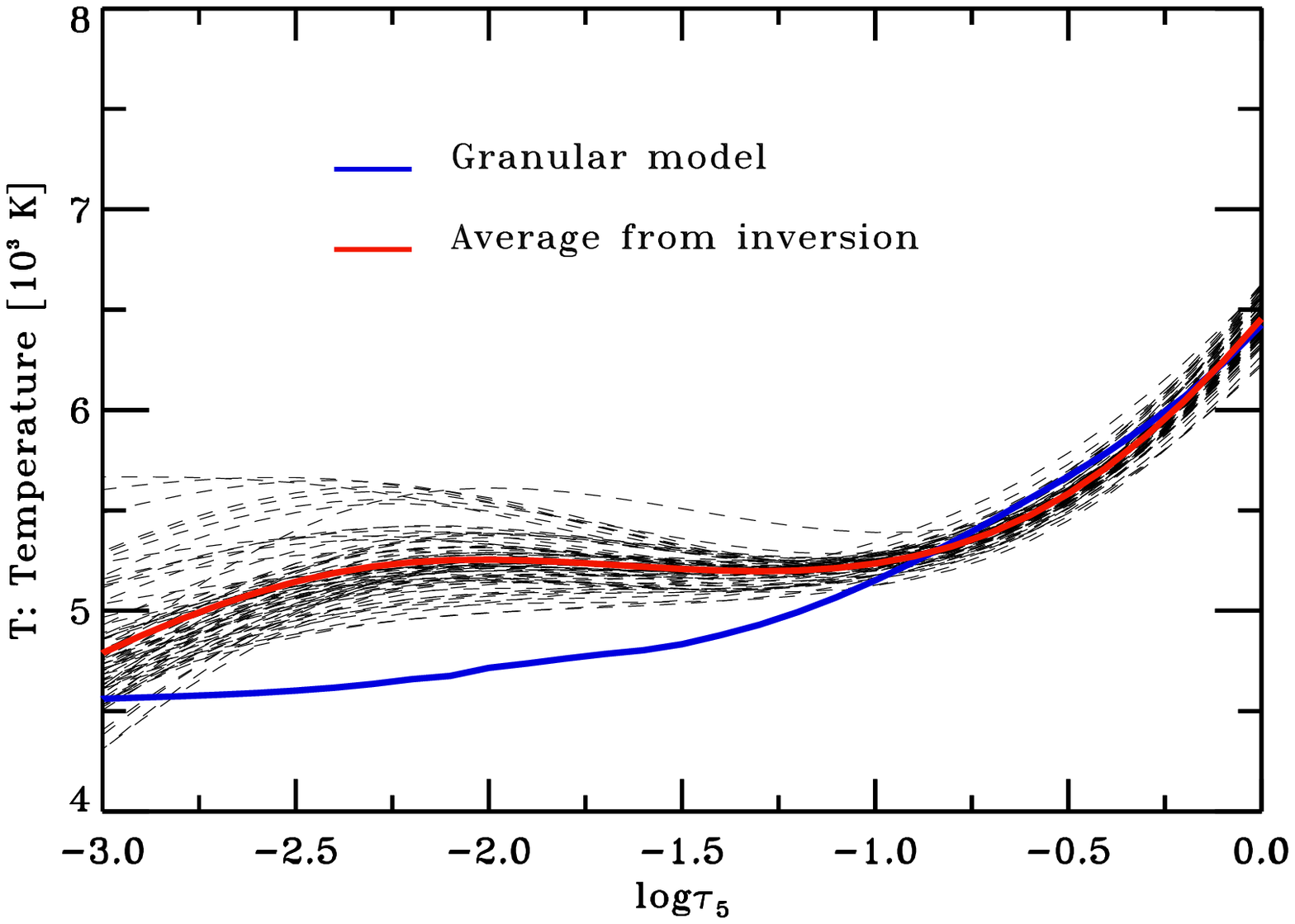} \\
\includegraphics[width=9cm]{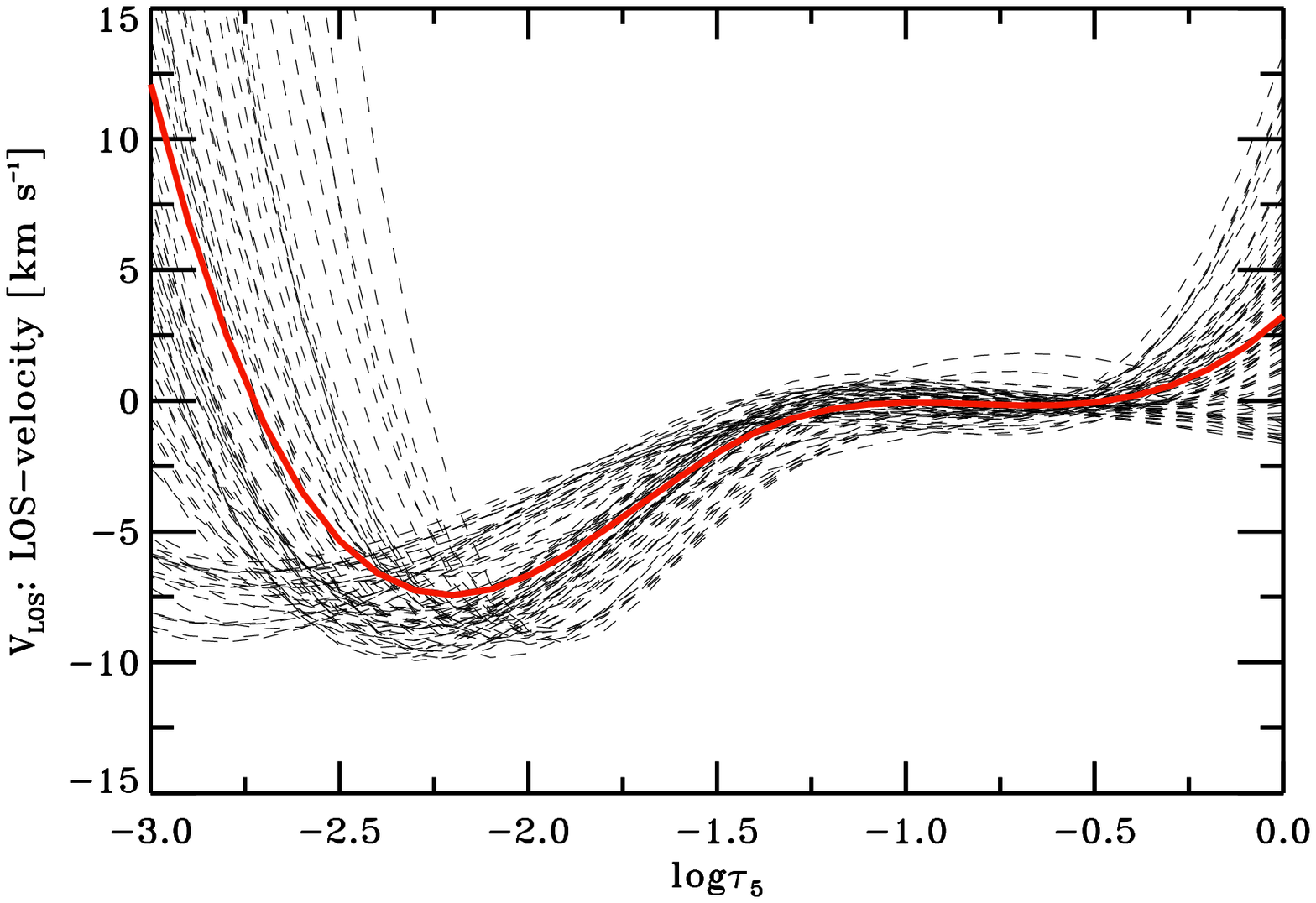} &
\includegraphics[width=9cm]{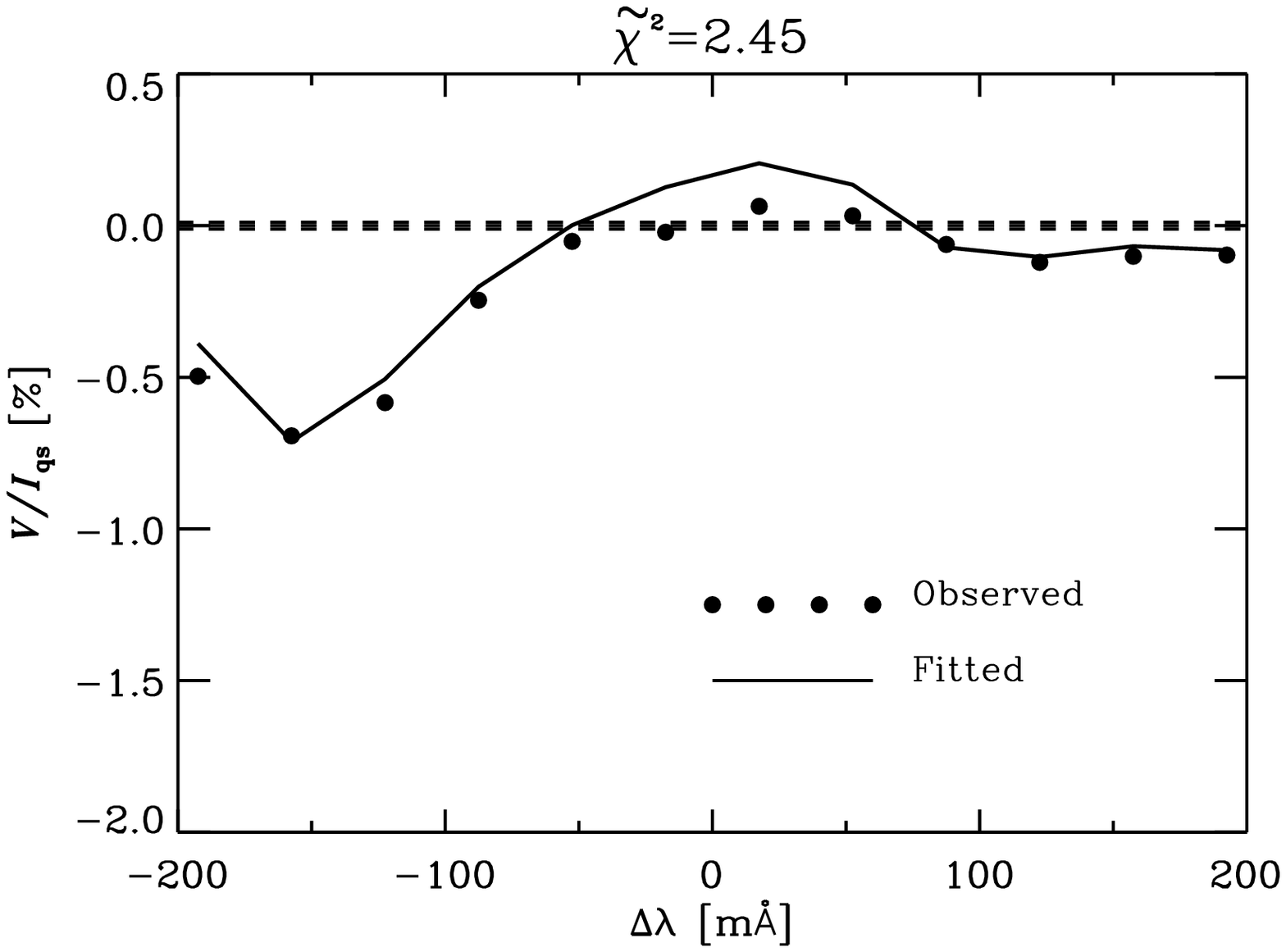} \\
\end{tabular}
\figcaption{{\it Top-left panel}: line-of-sight component of the magnetic field vector as a function of the optical depth $B_\parallel(\tau_5)$. 
{\it Top-right panel}: temperature as a function of the optical depth $T(\tau_5)$. {\it Bottom-left panel}: line-of-sight velocity as a function
of the optical depth $V_{\rm los}(\tau_5)$. Dashed-black lines show the results from the inversion of the Stokes profiles $I$ and $V$ from
each of the 123 selected pixels that belong to {\it family 1}. The red-solid line shows the average obtained from the individual results. 
Blue-solid line in the top-right panel corresponds to the temperature stratification in the granular model by Borrero \& Bellot Rubio (2002).
{\it Bottom-right panel}: average of the observed (circles) and of the fitted (solid line) Stokes $V$ profiles in all pixels belonging to {\it family} 1.
The mean value of $\chi^2$ from all the individual inversions is also indicated.}
\end{center}
\end{figure*}

\subsection{Family 2: polarity change at $\log\tau_5 \approx -2$.}

This family contains 434 of the 857 selected pixels (50.7 \%). As imposed by the classification criterion, $B_\parallel$
changes sign at around $\log\tau_5 \approx -2$ (Fig.~5; top-left panel). Similarly to {\it family 1}, the temperature in {\it family 2} also 
shows enhancements ($150-200$ K) compared to an average granule. Although not as large as in the first case, the increase occurs over 
all optical depths (Fig.~5; top-right panel). The line-of-sight velocities are again large, although in this case they always 
involve upflows (Fig.~5; bottom-left panel). These upflows are visible both in the upper-photosphere ($V_{\rm LOS} \approx -2$\kms~
 at $\log\tau_5 \approx -3$), and in the deep-photosphere ($V_{\rm LOS} \approx -7$\kms~ at $\log\tau_5 \approx 0$). Again, 
the line-of-sight velocity remains close to zero where the magnetic field changes polarity ($\log\tau_5 \approx -2$).\\

\begin{figure*}
\begin{center}
\begin{tabular}{cc}
\includegraphics[width=9cm]{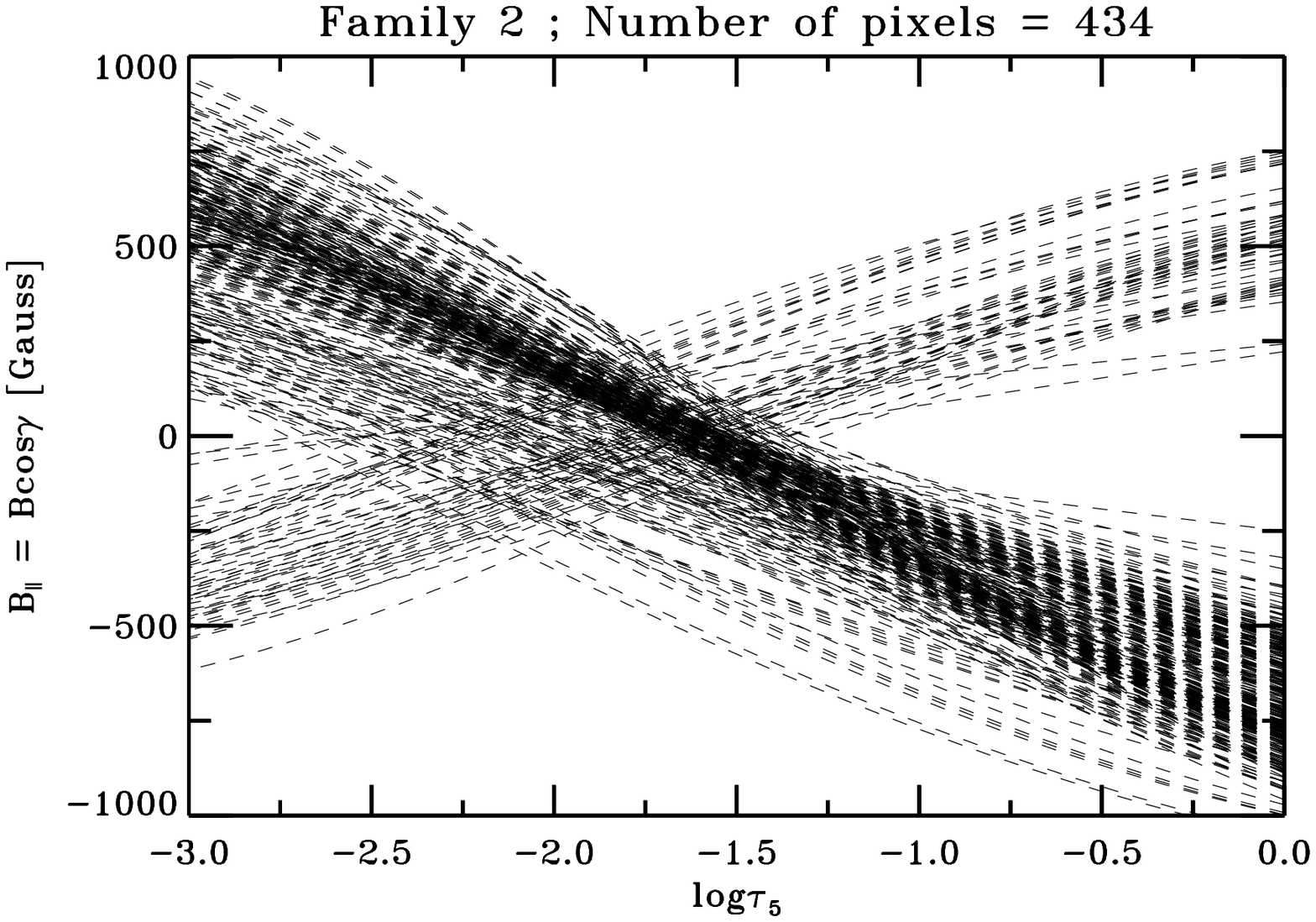} &
\includegraphics[width=9cm]{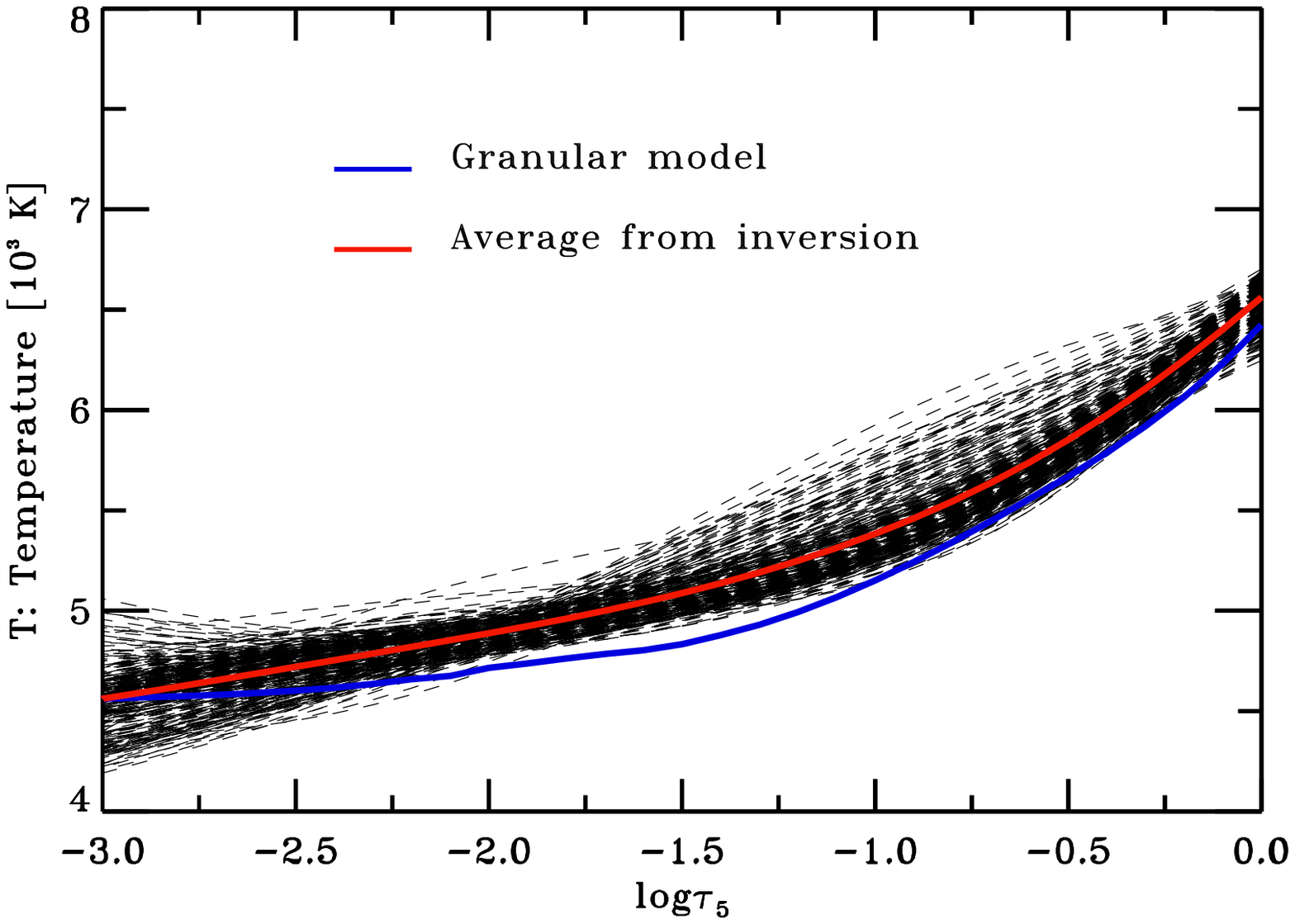} \\
\includegraphics[width=9cm]{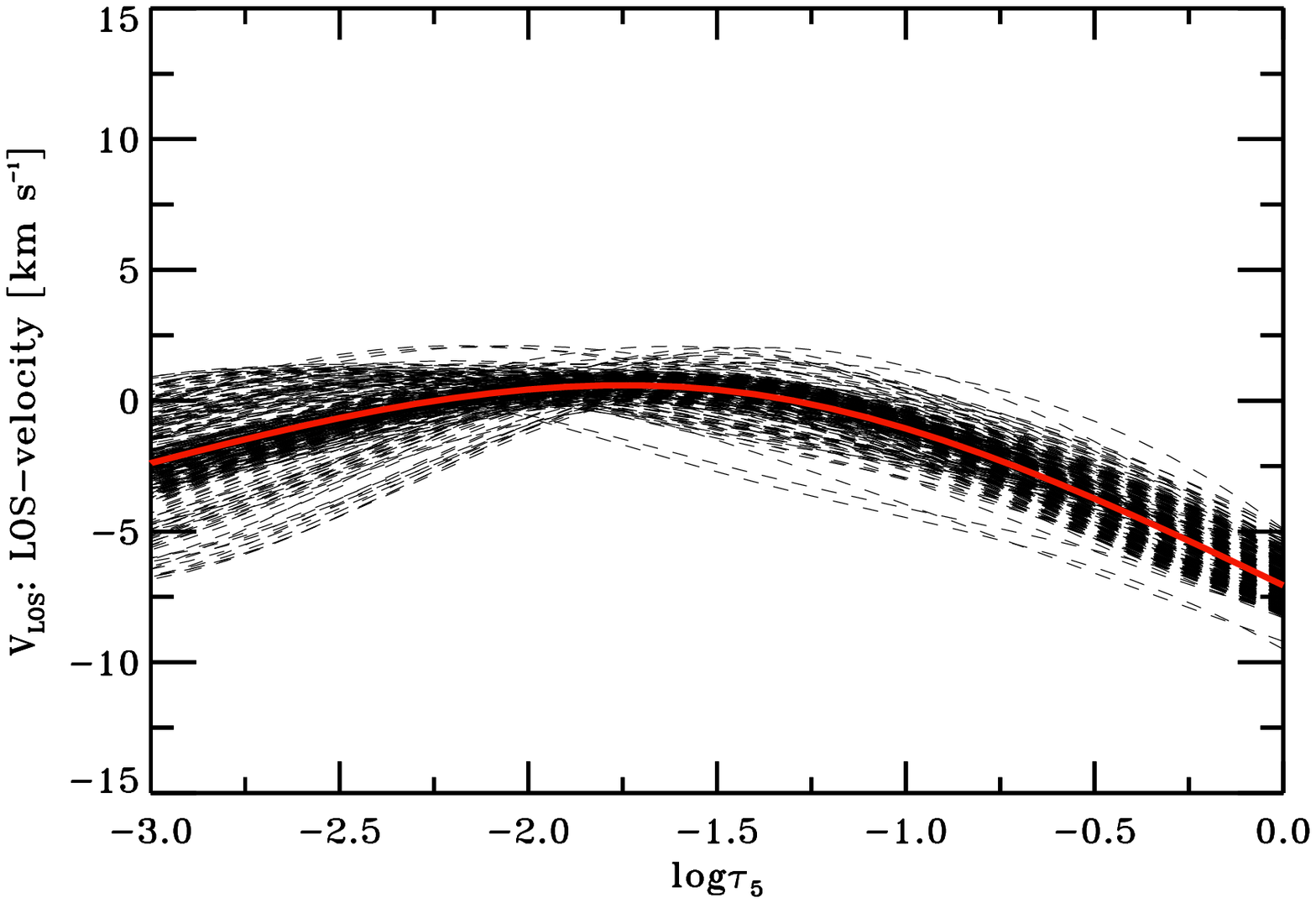} &
\includegraphics[width=9cm]{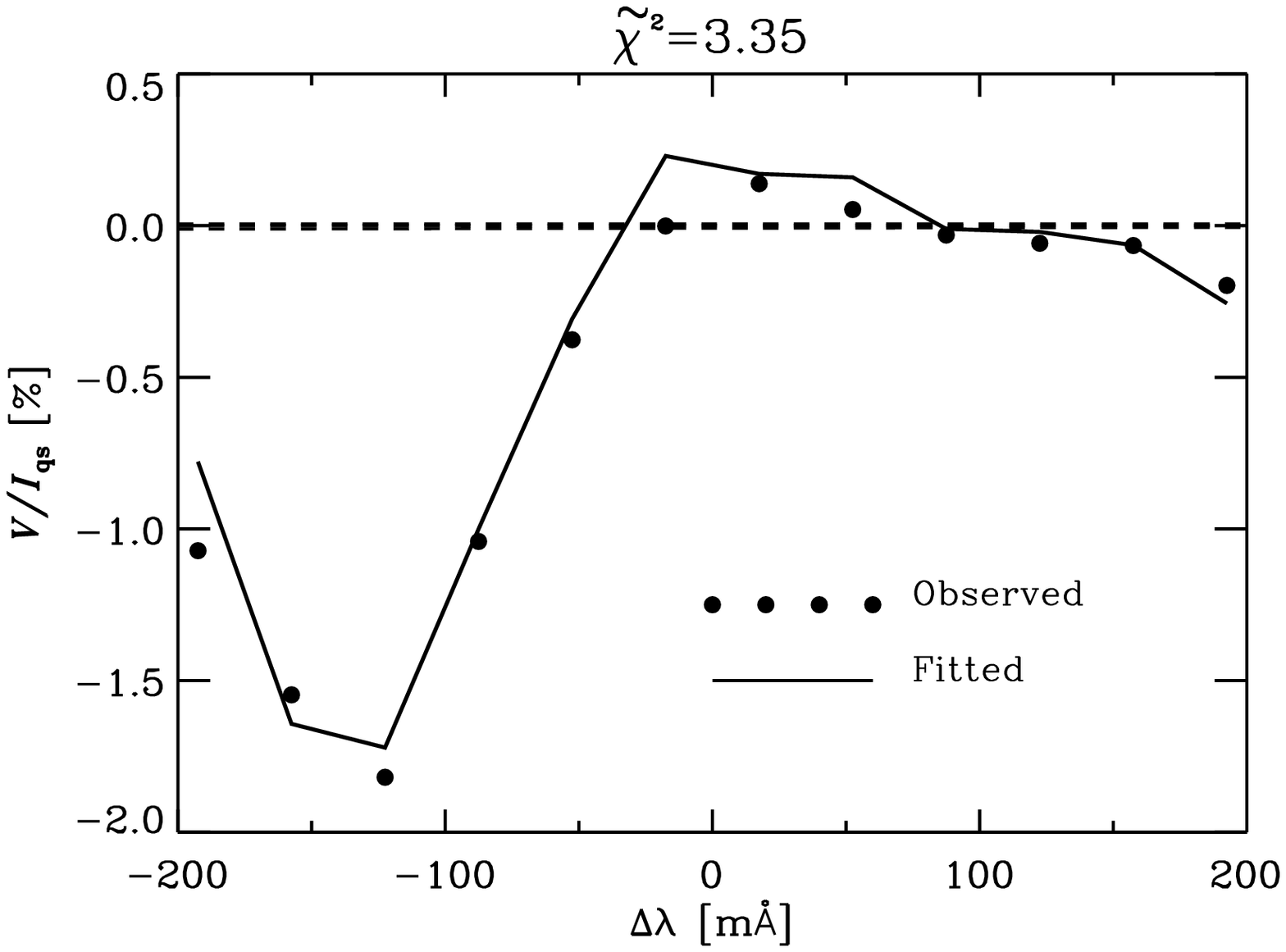} \\
\end{tabular}
\figcaption{Same as Figure 4 but showing the 434 pixels belonging to {\it family 2}.}
\end{center}
\end{figure*}

\subsection{Family 3: polarity change $\log\tau_5 < -3$ ?}

There are 300 pixels in this family, which corresponds to 35.0 \% of the total number. As our selection criterion
imposes, there is no real change in the polarity of the magnetic field vector (Fig.~6; top-left panel). Interestingly, $B_\parallel$ decreases
as $\log\tau_5$ decreases. If the decreasing trend continues towards higher photospheric layers the polarity would eventually switch, 
although that would happen close to the temperature minimum. As it happened in the two previous families, the temperature
in the mid-photosphere ($\log\tau_5 \in [-1,-2]$) is enhanced with respect to the average temperature of a granule (Fig.~6; top-right panel).
In this case, the enhancement is the largest ($\simeq 1000$ K) of the three studied families. Finally, 
the line-of-sight velocity changes from large upflows in the mid-photosphere, $V_{\rm LOS} \approx -8$\kms~ at $\log\tau_5 \approx -2$ 
(Fig.~6; bottom-left panel), to extreme downflows in the low-photosphere, $V_{\rm los} \approx 15-20$\kms~
at $\log\tau_5 \approx 0$. As in all previous families, the line-of-sight velocity drops to zero as the line-of-sight 
component of the magnetic field vector vanishes, which now happens in the upper-photosphere ($V_{\rm LOS} \rightarrow 0$ at 
$\log\tau_5 \approx -3$).\\

\begin{figure*}
\begin{center}
\begin{tabular}{cc}
\includegraphics[width=9cm]{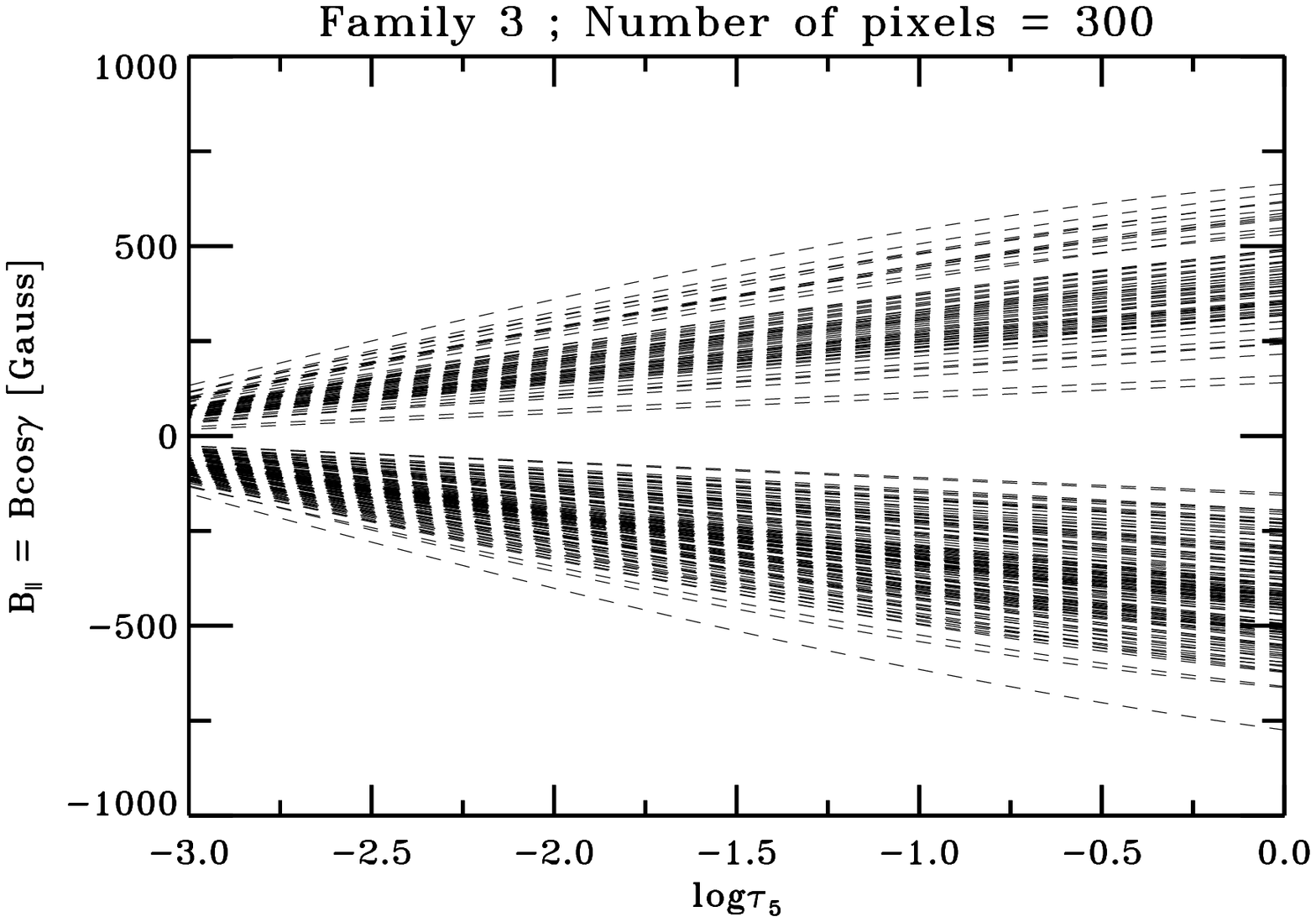} &
\includegraphics[width=9cm]{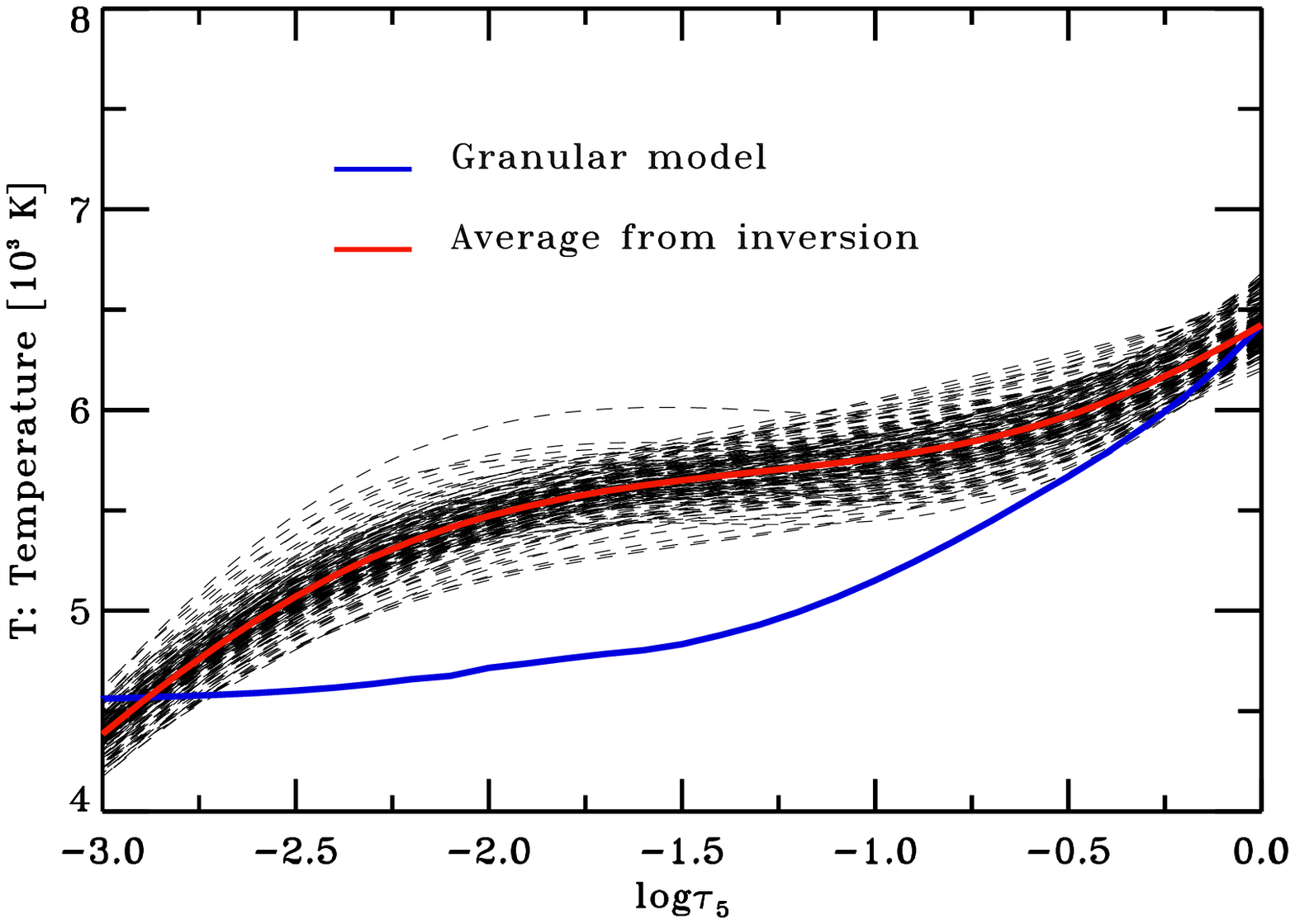} \\
\includegraphics[width=9cm]{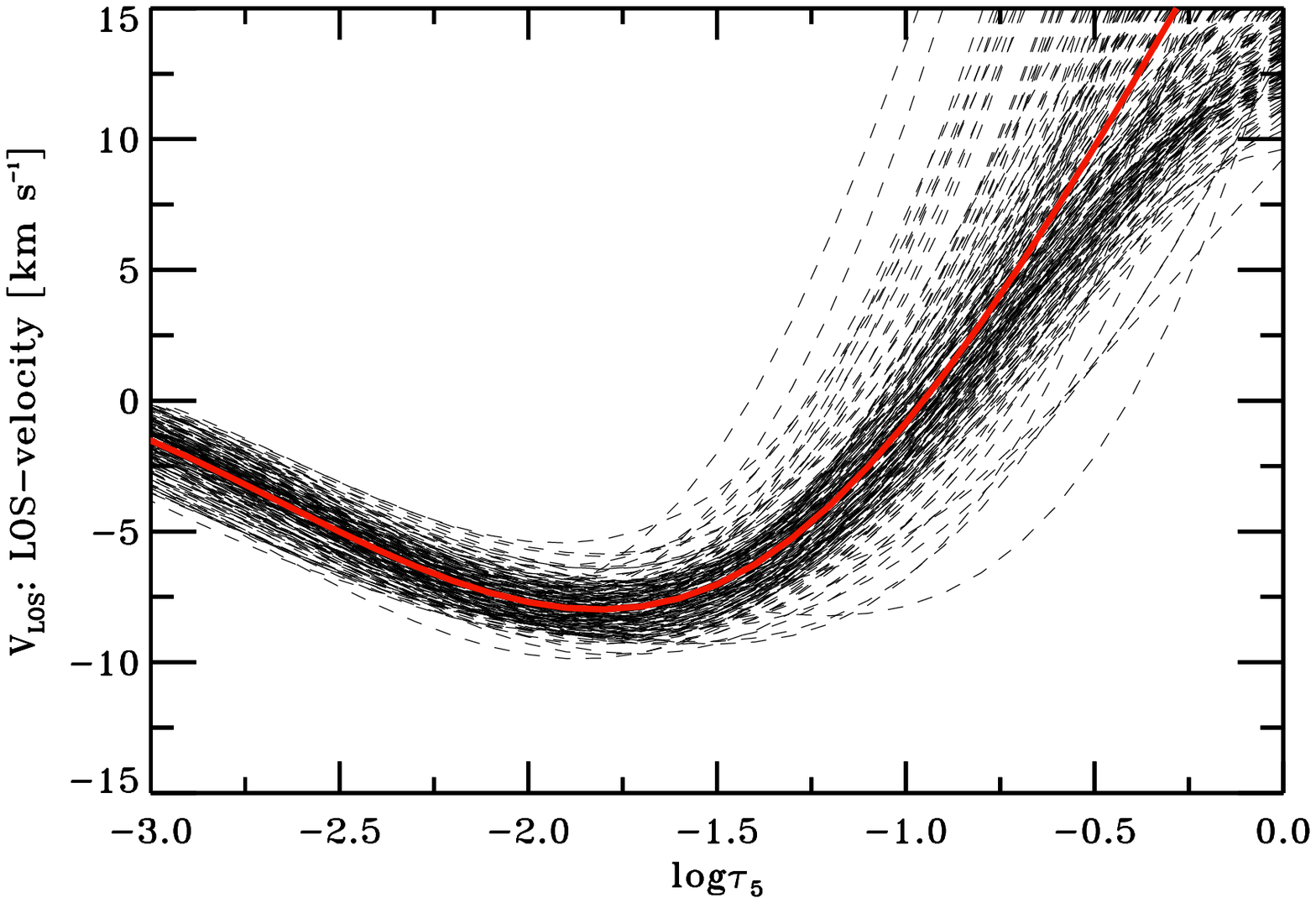} &
\includegraphics[width=9cm]{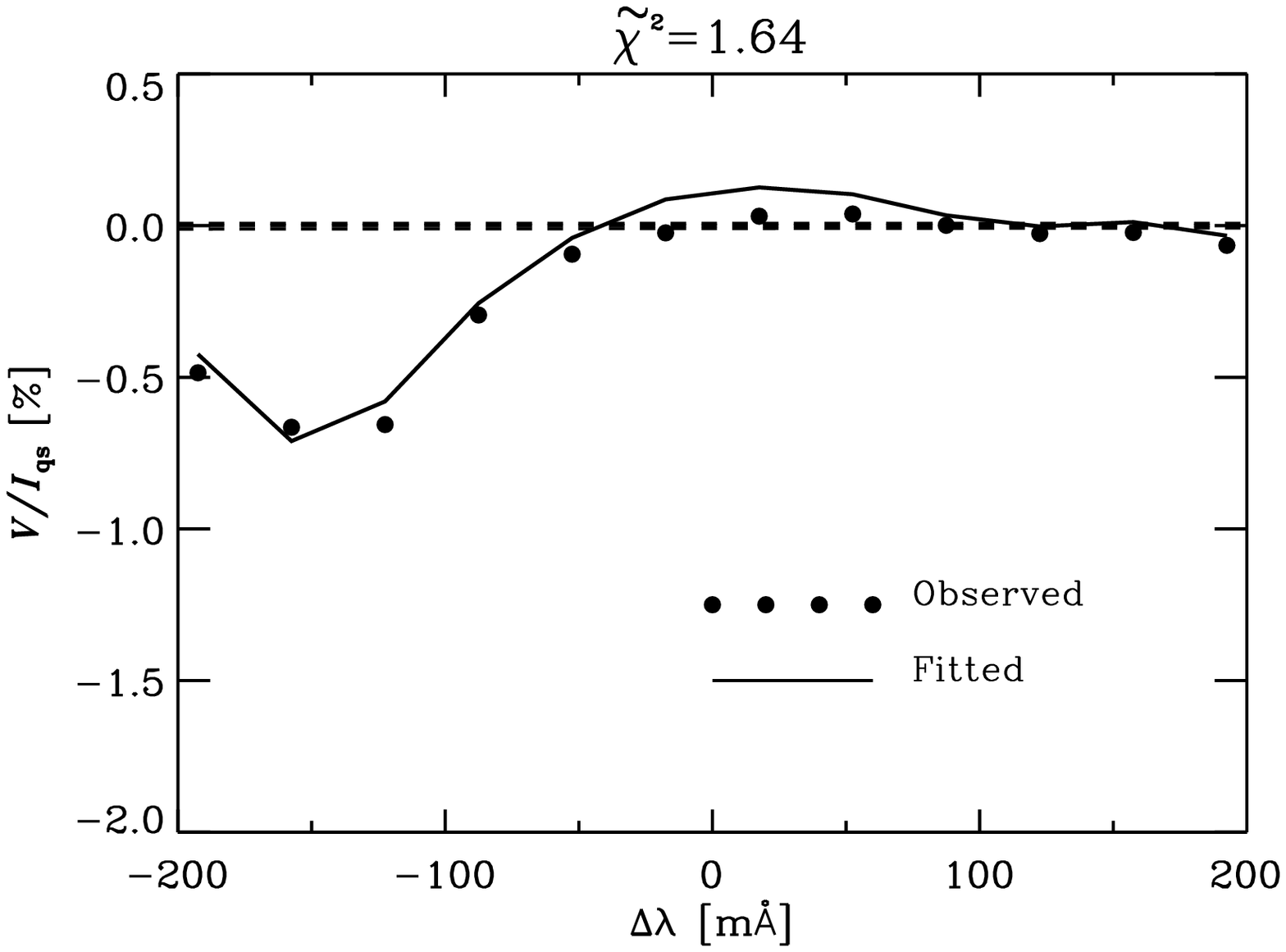} \\
\end{tabular}
\figcaption{Same as Figure 4 but showing the 300 pixels belonging to {\it family 3}.}
\end{center}
\end{figure*}

\section{Discussion on model choice}%

In this section we discuss the choice of model and free parameters in the inversion described
in Section 4. To make this choice we look at the observed Stokes profiles and try to establish
the most suitable model to fit them. The first step is to realize that, as illustrated in Fig.~3, 
all 857 selected Stokes $V$ profiles feature the almost complete lack of one of the lobes (produced by 
the $\Delta M = \pm 1$ transitions in the Zeeman pattern) in the circular polarization.\\

The question is now to decide whether these kind of profiles are really asymmetric, or on the other hand, 
they are symmetric but the missing lobe in Stokes $V$ is located at $\Delta \lambda < -200$ m{\AA}, hence 
lying just outside of the region scanned by IMaX. This question can be answered by looking at the wavelength 
region around $\Delta \lambda \approx 150-200$ m{\AA} in Figs. 4, 5 and 6 (bottom-right panels). The average 
observed Stokes profiles (filled circles) for all families around this region is negative and, although small, 
clearly above the noise level. We have established that this corresponds to the Stokes $V$ signal from the 
nearby \ion{Fe}{1} 5250.653 {\AA} spectral line by repeating all the inversions described in Section 4, 
but excluding this spectral line (see Table 1). In this case, the fitted
profiles in the bottom-right panels in Figs.~4-5-6 (solid-black lines) fail to reproduce the negative values of
Stokes $V$ around $\Delta \lambda \approx 150-200$ m{\AA}. With this in mind we now move to $\Delta \lambda < -200$
m{\AA} and conclude that, if there is a missing positive lobe in Stokes $V$ there, then the signal from 
\ion{Fe}{1} 5250.653 {\AA} at $\Delta \lambda \approx 150-200$ should also be positive. However, it is negative, 
and therefore we conclude that the observed Stokes $V$ profile from \ion{Fe}{1} 5250.653 {\AA} are indeed 
asymmetric and that there is no missing positive lobe at $\Delta \lambda < -200$ m{\AA}.\\

Once established that the observed circular polarization profiles are highly asymmetric, we follow Solanki \& 
Montavon (1993) and Landolfi \& Landi Degl'Innocenti (1996) and consider that those asymmetries are caused
by the simultaneous effect of gradients in the line-of-sight velocity $V_{\rm LOS}$, and in line-of-sight component of 
the magnetic field vector $B_\parallel = B \cos\gamma$. The large number of nodes allowed in $V_{\rm LOS}(\tau_5)$ 
(see Sect.~4) is a direct consequence of the need to fit these extremely asymmetric Stokes $V$ profiles.\\

Gradients in $B_\parallel$ have been included by, as already mentioned, allowing two nodes in $\gamma(\tau_5)$ and one node 
in $B(\tau_5)$. This combination is more general than allowing two nodes for $B(\tau_5)$ and only one for $\gamma(\tau_5)$. 
This is a consequence of the modulus of the magnetic field vector being defined as a positive quantity which makes, unlike 
the former case, the inversion with only one node in $\gamma(\tau_5)$ unable to yield solutions where $B_\parallel$ changes 
sign with optical depth. The possibility of obtaining solutions where $B_\parallel$ changes
sign does not imply that all inferred stratification will actually have it (i.e. {\it family} 3; see Sect.~5.3 and Fig.~6).\\

Since our discussion in the next section will rely heavily on the presence of a reversal in the polarity of the magnetic field, it is
crucial to establish whether this feature is really needed to reproduce the observed profiles. To this end we once more repeated 
our inversions (as described in Sect.~4) but employing this time one node in $\gamma(\tau_5)$ and two in $B(\tau_5)$. In this case, all retrieved 
stratification in $B_{\parallel}(\tau_5)$ show the same sign at all optical depths (as in {\it family 3}). Interestingly, the quality 
of the fits worsens significantly. The average value of the merit function, $\tilde{\chi}^2$, doubles in those pixels that belonged to {\it families} 
1 and 2. Meanwhile, $\tilde{\chi}^2$ also increases in those pixels that belonged to {\it family 3}, but comparatively less ($< 50$ \%). 
These results suggest that allowing two nodes in $\gamma(\tau_5)$ is necessary to successfully fitting the observed 
Stokes profiles, and therefore also that, the inferred reversal in the polarity of the magnetic field $B_\parallel$ (in {\it families} 1 an 2) 
is not an artifact imposed by our choice of model, but rather a characteristic feature of these events.\\

Finally, as previously discussed, a model that contains only one component assumes that the solar photophere is 
laterally homogeneous within each observed pixel, or at least, assumes that the vertical variations in the physical parameters 
play a more important role than the horizontal ones in the formation of the observed spectral line. While there is no 
guarantee that this is indeed the case, the high spatial resolution achieved by {\sc Sunrise}/IMaX makes this approximation a 
reasonable first step in the study of these events.\\

\section{Discussion and conclusions}%

The results presented here confirm our initial conclusions in our previous studies (Borrero et al. 2010, 2012) 
on these extremely shifted polarization signals. Namely that, {\bf a)} they occur mostly at the center or edges of granular cells; {\bf b)} they
are characterized by supersonic upward velocities; {\bf c)} they involve magnetized plasma, and d) magnetic fields of opposity
polarities are oftentimes ($\simeq 70$ \% of the cases) seen in their proximity ($\simeq 2"$). In addition to this,
the inversion of the Stokes profiles reveals that these events seem to belong to three distinct families. These families frequently 
present features such: {\bf e)} temperature enhancement of a few hundred Kelvin in the mid-photosphere; {\bf f)} shift from supersonic 
upflows to supersonic downflows at some height in the photosphere; and {\bf g)} presence of a reversal in the polarity of the magnetic 
field vector also at some height in the photosphere at the exact location where the event occurs.\\

Owing to their common features, and under the assumption that only one physical mechanism is responsible for all the observed events, 
it would be almost straightforward to consider magnetic reconnection as their probable cause: magnetic field lines of opposite polarity 
coalesce and the energy stored in the magnetic field is released into kinetic and thermal energy. The two different polarities would 
channel the plasma in different directions giving rise to both positive and negative line-of-sight 
velocities (Rezaei et al. 2007; Cameron et al. 2011). Unfortunately, not all investigated pixels share the aforementioned properties. For instance, only some 
cases ({\it families} 1 and 3) show both positive and negative supersonic line-of-sight velocities 
$V_{\rm LOS}$, while {\it family} 2 posseses only $V_{\rm LOS} < 0$ (upflows). In addition, the reversal in the polarity of the magnetic 
field vector is not always present (e.g {\it family} 3; see Fig.~6), and only in the case of {\it family 1} (14.3 \% of the cases) 
the reversal in $B_\parallel$ occurs at the same location as the change in the sign of $V_{\rm LOS}$.\\

On the one hand, taking into account that the interaction between magnetic fields and granular convection leads to a rich variety of 
phenomena, it is conceivable that the differences between the inferred families are caused by the underlying physical mechanism being 
different in each case. Although there are many possible candidates, a search across the available literature 
(Steiner et al 1998; Cheung et al. 2008; and references therein) does not reveal any mechanism that reproduces the observational 
features, neither in general nor of the individual families, of the events studied in this work. For instance, the supersonic flows predicted
by Cattaneo et al. (1990) and later observed by  Ryb\'ack et al. (2004) and Bellot Rubio (2009),
occur above granules and involve supersonic flows, but they are mostly horizontal and therefore their contribution
to $V_{\rm LOS}$ is unlikely to be large. Flux-emergence processes described in Cheung et al. (2008) take place
also in granules, but they do not seem to involve very large upflows. {\it Swaying motions} in flux tubes 
(Steiner et al. 1998) excite up-ward propagating shock fronts, but they occur mainly above intergranular lanes.
Finally, {\it vortex tubes} that were originally found in granules (Steiner et al. 2010) have been recently associated with
fast upflows but on nearby dark lanes (Yurchyshyn et al. 2011) and therefore, they probably correspond to a different kind of event.\\

On the other hand, one could attempt to salvage the hypothesis of reconnection by adopting different views. For instance, we could argue that 
one cannot expect all families to be fully consistent with the classic picture of magnetic reconnection, because they might correspond 
to different stages in the temporal evolution of the events (see Cameron et al. 2011). In order to rule out or to confirm 
this possibility, one would need an uninterrupted, and possibly longer, time-series of L12-2 data. Hopefully, this will 
be possible in the incoming second flight from {\sc Sunrise}/IMaX that is scheduled to take place in the summer of 2013. It can also 
be argued that, even if events belonging to {\it family} 3 do not show a polarity reversal in the magnetic field, this reversal 
can indeed take place in the upper-photosphere ($\log\tau_5 < -3$; see Sect.~5.3). Moreover, even if the polarity reversal is not present 
on the same pixel, in Borrero et al. (2010) we had already detected opposite polarities within 2" in 70\% of the events. In the future,
it would be very interesting to combine IMaX observations with data from the upcoming IRIS mission, to study a possible relationship
between these reconnection events and the presence of mostly unipolar regions (coronal holes) and/or type II spicules in the Chromosphere
(McIntosh et al. 2011).\\

\begin{acknowledgements}
Comments from Oskar Steiner and Rolf Schlichenmaier are gratefully acknowledged. Thanks to Fatima Rubio for 
providing the heliocentric angle of the observations, and to Tino Riethm\"uller for pointing out an error in
the identification of the \ion{Co}{1} line spectral line in Figure 1. The German contribution to {\sc Sunrise} is 
funded by the Bundesministerium f\"{u}r Wirtschaft und Technologie through Deutsches Zentrum f\"{u}r 
Luft-und Raumfahrt e.V. (DLR), Grant No. 50~OU~0401, and by the Innovationsfond of the President 
of the Max Planck Society (MPG). The Spanish contribution has been funded by the Spanish MICINN 
under projects ESP2006-13030-C06 and AYA2009-14105-C06 (including European FEDER funds).
\end{acknowledgements}

\end{document}